\documentclass[useAMS,usenatbib]{mn2e}
\usepackage{graphicx, txfonts, natbib}
\usepackage{epsf}
\usepackage{amssymb}
\usepackage{epstopdf}
\usepackage{longtable}
\usepackage{subfigure}
\usepackage{comment}
\usepackage[colorlinks=false,dvips]{hyperref}
\newcommand{\msun}{\mbox{M$_{\odot}$}}

\newcommand{\kms}{\mbox{$\rm{km}\,s^{-1}$}}

\DeclareMathAlphabet{\mathsc}{OT1}{cmr}{m}{sc}
\def\testbx{bx}%
\DeclareRobustCommand{\ion}[2]{%
\relax\ifmmode
\ifx\testbx\f@series
{\mathbf{#1\,\mathsc{#2}}}\else
{\mathrm{#1\,\mathsc{#2}}}\fi
\else\textup{#1\,{\mdseries\textsc{#2}}}%
\fi}

\newcommand{\Nai}{\ion{Na}{i}}

\newcommand{\Feii} {\ion{Fe}{ii}}
\newcommand{\FeiiF} {[\ion{Fe}{ii}]}
\newcommand{\Caii} {\ion{Ca}{ii}}
\newcommand{\CaiiF} {[\ion{Ca}{ii}]}

\newcommand{\FeiiiF} {[\ion{Fe}{iii}]}

\newcommand{\Nii} {[\ion{N}{ii}]}

\newcommand{\SiII} {\ion{Si}{ii}}

\newcommand{\CoiiF}{[\ion{Co}{ii}]}
\newcommand{\CoiiiF}{[\ion{Co}{iii}]}

\newcommand{\NiII}{\ion{Ni}{ii}}
\newcommand{\NiIIF}{[\ion{Ni}{ii}]}
\newcommand{\Feiii} {\ion{Fe}{iii}}
\newcommand{\FeiF} {[\ion{Fe}{i}]}

\begin{document}
\title[SN Ia explosion parameters] {Using late-time optical and near-infrared spectra to constrain Type Ia supernova explosion properties}
 \author[K. Maguire et al.]
 {K.~Maguire,$^1$\thanks{E-mail: kate.maguire@qub.ac.uk } S.~A.~Sim,$^1$, L.~Shingles,$^1$ J.~Spyromilio,$^{2}$ A.~Jerkstrand,$^{3}$  M.~Sullivan,$^{4}$ 
 \newauthor  T.-W.~Chen,$^{5}$  R.~Cartier,$^{6}$ G.~Dimitriadis,$^{4,7}$ C.~Frohmaier,$^{8}$ L.~Galbany,$^{9}$ C. P. Guti{\'e}rrez,$^{4}$
 \newauthor  G.~Hosseinzadeh,$^{10,11}$ D.~A.~Howell,$^{10,11}$ C.~Inserra,$^{4}$ R.~Rudy,$^{12}$ J.~Sollerman$^{13}$ \\
      $^1$Astrophysics Research Centre, School of Mathematics and Physics, Queen's University Belfast, Belfast BT7 1NN, UK\\
       $^2$European Southern Observatory, Karl-Schwarzschild Strasse 2, D-85748 Garching bei M\"unchen, Germany\\
       $^3$Max-Planck Institut f\"ur Astrophysik, Karl-Schwarzschild Strasse 1, D-85748 Garching bei M\"unchen, Germany \\
      $^4$Department of Physics and Astronomy, University of Southampton, Southampton, SO17 1BJ, UK\\
      $^{5}$Max-Planck-Institut f{\"u}r Extraterrestrische Physik, Giessenbachstrasse 1, D-85748, Garching bei M\"unchen, Germany \\
      $^{6}$Cerro Tololo Inter-American Observatory, National Optical Astronomy Observatory, Casilla 603, La Serena, Chile\\  
       $^{7}$Department of Astronomy and Astrophysics, University of California, Santa Cruz, CA 95064, USA\\
       $^{8}$Institute of Cosmology and Gravitation, University of Portsmouth, Portsmouth, PO1 3FX, UK\\
       $^9$PITT PACC, Department of Physics and Astronomy, University of Pittsburgh, Pittsburgh, PA 15260, USA\\
        $^{10}$Las Cumbres Observatory, 6740 Cortona Dr Ste 102, Goleta, CA 93117-5575, USA\\
        $^{11}$Department of Physics, University of California, Santa Barbara, CA 93106-9530, USA\\
 $^{12}$Space Science Applications Laboratory, The Aerospace Corporation, P.O. Box 92957, Los Angeles, CA 90009, USA \\
 $^{13}$The Oskar Klein Centre, Department of Astronomy, Stockholm University, AlbaNova, SE-10691 Stockholm, Sweden \\
 }
\maketitle

\begin{abstract}
The late-time spectra of Type Ia supernovae (SNe Ia) are powerful probes of the underlying physics of their explosions. We investigate the late-time optical and near-infrared spectra of seven SNe Ia obtained at the VLT with XShooter at $>$200 d after explosion. At these epochs, the inner Fe-rich ejecta can be studied. We use a line-fitting analysis to determine the relative line fluxes, velocity shifts, and line widths of prominent features contributing to the spectra (\FeiiF, \NiIIF, and \CoiiiF). By focussing on \FeiiF\ and \NiIIF\ emission lines in the $\sim$7000--7500 \AA\ region of the spectrum, we find that the ratio of stable \NiIIF\ to mainly radioactively-produced \FeiiF\ for most SNe Ia in the sample is consistent with Chandrasekhar-mass delayed-detonation explosion models, as well as sub-Chandrasekhar mass explosions that have metallicity values above solar. The mean measured Ni/Fe abundance of our sample is consistent with the solar value. The more highly ionised \CoiiiF\ emission lines are found to be more centrally located in the ejecta and have broader lines than the \FeiiF\ and \NiIIF\ features. Our analysis also strengthens previous results that SNe Ia with higher \SiII\ velocities at maximum light preferentially display blueshifted \FeiiF\ 7155 \AA\ lines at late times. Our combined results lead us to speculate that the majority of normal SN Ia explosions produce ejecta distributions that deviate significantly from spherical symmetry.
\end{abstract}


\section{Introduction} \label{intro}
Type Ia supernovae (SNe Ia) are the violent luminous deaths of carbon-oxygen white dwarfs in binary systems. Despite being precision probes of the cosmological parameters \citep[e.g.][]{2014A&A...568A..22B}, we still do not understand the mechanisms that lead to their explosions. Proposed explosion scenarios are generally split into two broad classes depending on the properties of their companion stars at the time of explosion. These are the `double-degenerate' scenario \citep{1984ApJS...54..335I,1984ApJ...277..355W}, where the companion star is another white dwarf and the `single-degenerate' scenario \citep{1973ApJ...186.1007W}, where the companion star is a main sequence or giant star. Alternatively the competing scenarios can be framed in terms of the primary white dwarf mass at the time of explosion: those that explode near the Chandrasekhar mass limit, `M$_{ch}$ explosions' or those that explode significantly below the Chandrasekhar mass limit, `sub-M$_{ch}$ explosions'. Historically, M$_{ch}$ explosions were thought to be responsible for the bulk of normal SNe Ia. However, sub-M$_{ch}$ explosions have been the focus of a number of recent studies. These sub-M$_{ch}$ models include both double-detonation scenarios \citep{1982ApJ...253..798N,1990ApJ...361..244L,2007ApJ...662L..95B,2010A&A...514A..53F,2014ApJ...785...61S}, where a thin layer of He on the surface of the white dwarf ignites to cause a subsequent core detonation, as well as double white-dwarf mergers where the individual masses can be well below the Chandrasekhar mass \citep{2007MNRAS.380..933Y,2011ApJ...737...89D,2012ApJ...747L..10P}. 

Observational evidence in favour (and disfavour) of both M$_{ch}$ and sub-M$_{ch}$ explosions has been put forward, with the current evidence suggesting that there is more than one channel for producing normal SNe Ia.  This evidence comes from a number of probes, not limited to companion star interaction at early times, signatures of circumstellar material, presence of H and He features stripped from a companion star, studies of SN remnants nucleosynthetic yields  \citep[e.g.][]{2005A&A...443..649M,2007ApJ...670.1275L,2011Sci...333..856S,2013MNRAS.436..222M,2013A&A...559L...5S,2014MNRAS.tmp..560S,2015Natur.521..332O,2016ApJ...820...92M,2017ApJ...843...35M}.

A key distinguishing prediction between M$_{ch}$ and sub-M$_{ch}$ explosions is the different ratios of stable to radioactive isotopes of Fe-group elements produced. This ratio depends on the central density of the white dwarf at the time of explosion  \citep{2004ApJ...617.1258H}. Steadily-accreting M$_{ch}$ models predict significantly higher central densities ($\sim$10$^9$ g cm$^{-3}$) than sub-M$_{ch}$ models where the central densities are expected to be much lower  \citep[$\sim$10$^{7}$ g cm$^{-3}$;][]{2012ApJ...747L..10P,2009ApJ...705L.128R,2018ApJ...854...52S}. Higher central densities manifest themselves as larger ratios of stable (e.g.~$^{54}$Fe, $^{58}$Ni) to radioactively-produced ($^{56}$Fe) isotopes due to increased electron capture in these high-density environments.

Investigating the central density requires spectra at late times when the inner regions of the ejecta are visible. This occurs at phases of $\gtrsim$200 d after peak magnitude when the outer ejecta have become transparent and we can see the inner Fe-group dominated regions. Detailed modelling has been performed to provide a comparison with a number of nearby SNe Ia  \citep[e.g.][]{1980PhDT.........1A,1992ApJ...400..127R,1992MNRAS.258P..53S,1997ApJ...483L.107L,2007Sci...315..825M,2015MNRAS.450.2631M,2015ApJ...814L...2F,2017ApJ...845..176B,2018MNRAS.474.3187W}. However, there are many differences between the models that depend on the inputs to the code, as well as in the details of the radiative transfer codes themselves. A good match to the shape and strength of the Ni and Fe complex at $\sim$7300 \AA\ is difficult to achieve, which results in different implied masses of neutron-rich stable Ni material. Some predicted the presence of significant quantities of stable Ni for SN 2011fe \citep{2015MNRAS.450.2631M}, while others suggested for the same SN that stable Ni is not required \citep{2017ApJ...845..176B}. Therefore, we have adopted an agnostic approach to determining the abundance ratios using empirical, model-independent methods. However, we stress that radiative transfer modelling of SNe Ia at late-times is vital for providing detailed constraints on ejecta abundances and understanding SN Ia explosion physics.

The late-time spectra of SNe Ia have also provided information on potential asymmetries in the ejecta \citep{2006ApJ...652L.101M,2007ApJ...661..995G,2010ApJ...708.1703M}. A link between maximum light \SiII\ velocity gradients and the wavelength shifts of late-time emission lines was identified, with SNe Ia with high \SiII\ velocities having preferentially redshifted late-time emission lines \citep{2010Natur.466...82M}.  It was suggested  that variations in the observed properties (light-curve widths, line velocities) are a consequence of different viewing angles of an explosion with an off-centre ignition point.  A tentative link between the late-time velocity shifts and SN Ia maximum-light colours has also been studied \citep{2011A&A...534L..15C,2011MNRAS.413.3075M,Blo12}.

In this paper, we investigate the largest sample to date of optical through near-infrared (NIR) spectral observations of SNe Ia at phases later than $\sim$200 d. At these epochs, the inner Fe-rich regions of the ejecta can be used to investigate the extent of line-emitting regions, signatures of asymmetry, and the relative abundances of stable to radioactive elements. In Section \ref{sec:obs_data}, we present the VLT+XShooter observations, complementary maximum-light spectral and photometric observations, as well as details of a comparison literature sample. In Section \ref{sec:line_fitting}, the line-fitting analysis is described, combined with specific details of fitting for the prominent spectral features. The  measurements of the late-time velocity shifts,  the colour evolution, and calculations of the Ni to Fe abundance ratio are detailed in Sections \ref{sec:shifts}, \ref{sec:colour}, and \ref{sec:nife_calc}, respectively. A discussion of the results is presented in Section \ref{sec:discussion}, and the conclusions are given in Section \ref{sec:conc}.

\begin{table}
  \caption{Observational details for previously unpublished late-time spectra observed at the VLT+XShooter.}
 \label{tab:spec_info}
\begin{tabular}{@{}lccccccccccccccccccccccccccccc}
  \hline
  \hline
SN &Observation& MJD$^a$&Phase$^b$&Exposure\\
&date& &(d)&time (s)&\\
\hline
\hline
SN2013aa&20140421&56769.2&425&7480&\\
ASASSN14jg&20150919&57285.2&322&2900\\
ASASSN15be&20151021&57317.3&266&5800\\
PSNJ1149&20160202&57421.3&206&2900\\
ASASSN15hx&20160626&57566.1&427&2980\\
\hline
\end{tabular}
 \begin{flushleft}
$^a$MJD = Modified Julian date.\\
$^b$Phase of late-time spectrum calculated with respect to maximum light.\\
  \end{flushleft}
\end{table}

\begin{table*}
  \caption{SN light curve, spectral, and host galaxy information.}
 \label{tab:SN_info}
\begin{tabular}{@{}lccccccccccccccccccccccccccccc}
  \hline
  \hline
Name &Phase$^a$&Host galaxy&Helio. \textit{z}$^b$&E(\textit{B-V})$^c$&Stretch$^{d}$&Date of max.$^{e}$&\textit{B - V}  at max.&\SiII\ vel. at max.&LC Ref.$^f$ & Spec. ref.$^f$\\
&(d)&&&(mag)&&&(mag)&(\kms)\\
\hline
SN 2012cg&339&NGC 4424    &0.0015       &0.20$^g$&1.098$\pm$0.022&20120603&0.14$\pm$0.04&10420$\pm$100&1&1\\ 
SN 2012fr&357&NGC 1365      &0.00546     &0.018&1.135$\pm$0.010&20121112&0.02$\pm$0.03 &11928$\pm$100&2,3&4\\  
SN 2012ht&433&NGC 3447a&0.00355        &0.026&0.877$\pm$0.022&20130103&$-$0.04$\pm$0.02&11000$\pm200$&1&1\\ 
SN 2013aa&360,425&NGC 5643&0.003999&0.169&1.146$\pm$0.019&20130221&$-$0.05$\pm$0.01&10600$\pm$100&1&1\\ 
SN 2013cs&303&ESO 576- G 017&0.00924&0.082&0.963$\pm$0.024&20130526&0.03$\pm$0.05&12700$\pm$150&5&6\\
SN 2013ct&229&NGC 0428       &0.00384   &0.025&--&20130504&--&--&--&--\\
PSNJ11492548&206&NGC 3915&0.0056    &0.025&--&20150712&--&10386$\pm$100&--&4\\ 
ASASSN14jg*&322&PGC 128348&0.014827&0.014& 1.053$\pm$0.007$^h$&20141102&--&10928$\pm$200&7&4\\
ASASSN15be*&266&GALEX J0252$^i$&0.0219&0.017&--&20150129&--&11900$\pm$200&--&4\\
ASASSN15hx*&427&GALEX J1343$^i$&0.0084&0.043&--&20150427&--&10407$\pm$200&--&4\\ 
\hline
\end{tabular}
 \begin{flushleft}
 $^*$NIR region of the spectrum is too low S/N for a detailed analysis to be performed. Used only in comparison with literature sample for line shift analysis of \protect \cite{2010Natur.466...82M}.\\
 $^a$Phase of late-time spectrum calculated with respect to maximum light. \\
 $^b$Heliocentric redshifts are from the Nasa Extragalactic Database (NED), apart from ASASSN15hx, where the redshift is measured from narrow H$\alpha$ emission from a WiFES \citep{2016PASA...33...55C} spectrum on 20150708, and SN 2012cg from \cite{2006AJ....131..747C}.\\
 $^c$Galactic E(\textit{B-V}) values from \protect \cite{2011ApJ...737..103S}.\\
$^d$Stretch is a measure of the light curve width \protect \citep{2008ApJ...681..482C}. \\
 $^e$Maximum light estimated from a light curve where available or estimated from the early-time spectra if not.\\
 $^f$Sources for maximum-light photometry and \SiII\ 6355 \AA\ velocities: (1)  \cite{2013MNRAS.436..222M} (2) \cite{2015MNRAS.451.1973S}, (3) \cite{2014ApJ...797....5Z} (4) \cite{2013ApJ...770..108C}, (5) this paper, (6) \cite{2016PASA...33...55C}, (7) \cite{2017MNRAS.472.3437G}.\\ 
 $^g$Additional host galaxy extinction of E(\textit{B-V})=0.18 mag was found for SN 2012cg by \cite{2012ApJ...756L...7S}. The value quoted here is the combined Galactic and host galaxy E(\textit{B-V}). \\
 $^h$Converted from a $\Delta$m$_{15}$(B) value to a SiFTO stretch using the relation from \cite{2008ApJ...681..482C}. \\
 $^i$Full name of galaxies:  GALEX J0252 = GALEXASC J025245.83-341850.6, GALEX J1343 =  GALEXASC J134316.80-313318.2 \\ 
 \end{flushleft}
\end{table*}

\section{Observations and Data Reduction}
\label{sec:obs_data}
Our analysis is performed on a sample of 8 spectra of 7 SNe Ia with high signal-to-noise (S/N) optical and NIR spectral coverage at epochs greater than 200~d after maximum light.  We use six spectra from the VLT+XShooter SN Ia sample presented in \cite{2016MNRAS.457.3254M}, combined with two new XShooter SN Ia spectra. Spectra of three additional SNe Ia obtained with XShooter are also presented but do not have high enough S/N in the NIR to be analysed in detail.

\subsection{Late-time spectroscopy with XShooter}
The spectra in our sample were obtained using the VLT+XShooter, which is an echelle spectrograph with three arms (UVB, VIS, NIR) covering the wavelength range $\sim$3000--25000 \AA\   \citep{2011A&A...536A.105V}. We used a setup with slit widths of 0.8, 0.9, and 0.9 arcsec in the UVB, VIS and NIR arms, respectively. The spectra were reduced using the REFLEX pipeline to produce flux-calibrated one-dimensional spectra \citep{2010SPIE.7737E..56M,2013A&A...559A..96F} as described in \cite{2016MNRAS.457.3254M}. 

Five late-time spectra of SNe Ia are presented for the first time here, ASASSN14jg, ASASSN15be, PSNJ11492548-0507138 (hereafter PSNJ1149), ASASSN15hx, and a second epoch of SN 2013aa (Table \ref{tab:spec_info}). These spectra were reduced in the same manner as in \cite{2016MNRAS.457.3254M}, and flux calibrated using photometric measurements from the acquisition images. For PSNJ1149 and SN 2013aa, the acquisition image photometry was calibrated using the XShooter acquisition camera zero-points. The spectra of ASASSN15hx were flux calibrated to coeval photometry using \textit{g'} acquisition imaging calibrated by comparison to catalogue Pan-STARRS\footnote{http://archive.stsci.edu/panstarrs/search.php} magnitudes \citep{2016arXiv161205560C,2016arXiv161205242M}. Due to the faintness of ASASSN14jg and ASASSN15be in their acquisition images, these spectra remain without absolute flux-calibration. However, since we deal only with relative line fluxes this does not affect any of our conclusions. 

Telluric corrections have been applied to the spectra using the \textsc{molecfit} package \citep{2015A&A...576A..77S,2015A&A...576A..78K}. For the spectra of ASASSN15be and ASASSN15hx, the low S/N did not allow for accurate fitting of the telluric regions. The spectra were corrected to their rest wavelengths using their heliocentric velocities and for Galactic extinction using the Galactic $\textit{E(B-V)}$ values given in Table \ref{tab:SN_info}.

\begin{figure*}
\includegraphics[width=18.cm]{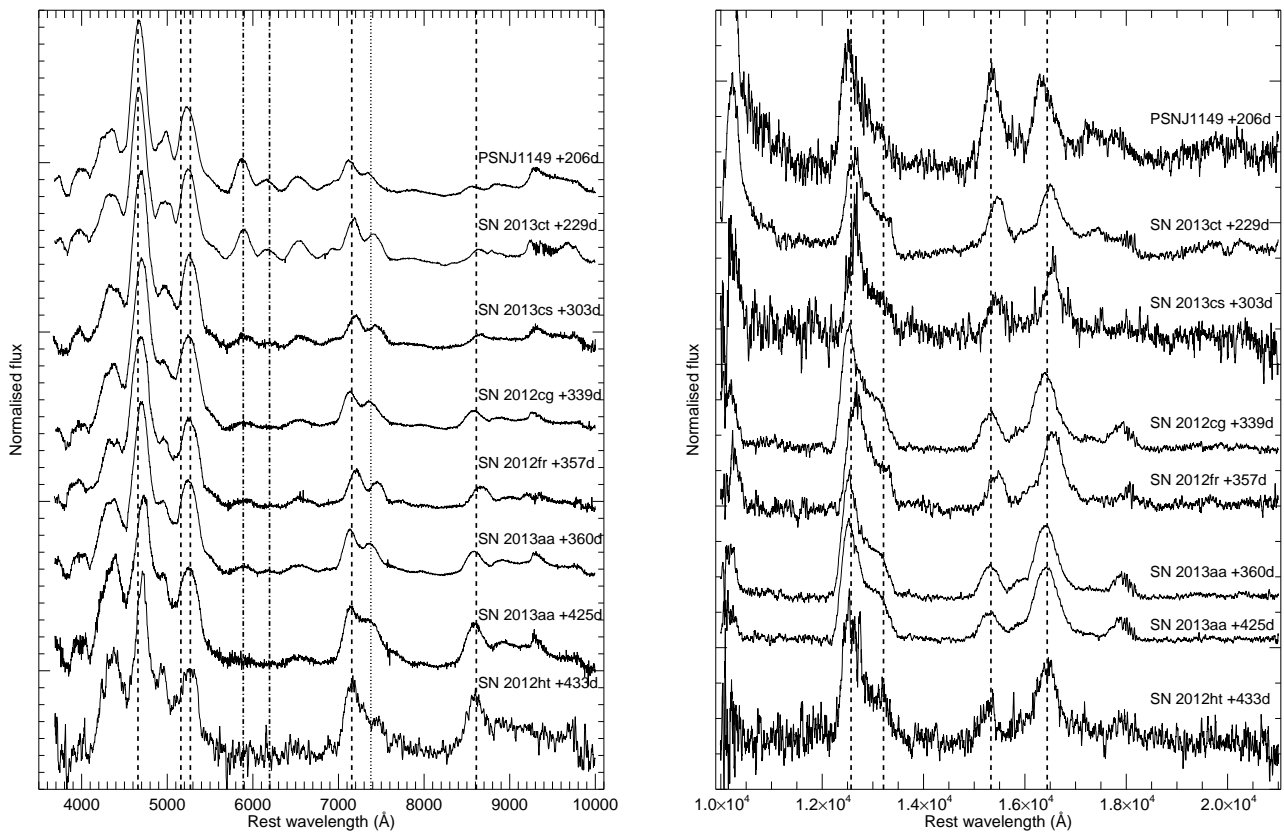} 
\caption{Optical (left) and NIR (right) spectral data of our full sample of SNe Ia at epochs greater than 200 d since maximum light. The rest wavelengths of prominent Fe-group lines are marked by the dashed lines  (from blue to red wavelengths: \FeiiiF\ 4658 \AA, \FeiiF\ 5159 \AA, \FeiiiF\ 5270 \AA, \FeiiF\ 7155 \AA, \FeiiF\ 8617 \AA, \FeiiF\ 12567 \AA, \FeiiF\ 13206 \AA, \FeiiF\ 15335 \AA\ and \FeiiF\ 16436 \AA), the \NiIIF\ 7378 \AA\ line by a dashed line, and \CoiiiF\ 5890, 6197 \AA\ by dash-dot lines.}
\label{optNIR_spec}
\end{figure*}

Of the five new late-time spectra, two (PSNJ1149, SN 2013aa) are included in the full analysis, while the spectra of ASASSN14jg, ASASSN15be and ASASSN15hx have lower S/N spectra, particularly in the NIR region of the spectrum. Therefore, they are excluded from the full analysis and only used in the comparison line-shift analysis presented in Section \ref{sec:shifts}. The optical and NIR spectra for the sample used in the detailed analysis are shown in Fig. \ref{optNIR_spec}.

\subsection{Maximum-light spectroscopy}
Maximum-light spectroscopy (within $\pm$5 d of max.) has been used to constrain the \SiII\ 6355 \AA\ velocities at maximum  for our late-time spectral sample. For SNe 2012cg, 2012fr, 2012ht, 2013aa and 2013cs, these values were obtained from the literature as detailed in Table \ref{tab:SN_info}. This is supplemented with unpublished data for four other SNe Ia in the sample, described below.

For PSNJ1149, a maximum-light classification spectrum was obtained with the Shane 3-m telescope of the Lick Observatory using the Visible and Near-Infrared Imaging Spectrograph (VNIRIS) \citep{rudyatel7825}, from which a \SiII\ velocity at maximum was measured.  ASASSN14jg was spectroscopically classified at the 2-m Las Cumbres Observatory (LCO) telescope \citep{2013PASP..125.1031B} with the FLOYDS spectrograph on 20141102 (at maximum) and from which a \SiII\ velocity was measured. For ASASSN15be, the Public ESO Spectroscopic Survey of Transient Objects \citep[PESSTO;][]{2015A&A...579A..40S} obtained a spectrum on 20150126 with EFOSC2 on ESO's New Technology Telescope. A \SiII\ velocity was measured from this spectrum that is consistent with a phase of $-$3 d with respect to maximum light.  ASASSN15hx was also spectroscopically observed at the 2-m LCO telescope with FLOYDS on 20150505 and 20150511 ($-$3 and $+$3 d from maximum). The \SiII\ velocity at maximum was taken as the mean of these two values. These \SiII\ 6355 \AA\ velocities are provided in Table \ref{tab:SN_info}.

\subsection{Photometry and light-curve analysis}
\label{phot}
The light curve properties (light-curve width, optical colour at maximum) for four SNe Ia in our sample were obtained from the literature (see references in Table \ref{tab:SN_info}). A \textit{gri}-band light curve of SN 2013cs obtained with the LCO 1-m telescope array is presented here for the first time (Table \ref{tab:sn13cs_phot}). The magnitude of the SN was measured using Point Spread Function photometry and calibrated using reference stars in the field of the SN with magnitudes from the Pan-STARRS Data Release 1 catalog\footnote{http://archive.stsci.edu/panstarrs/search.php}. The light curve width (`stretch') of SN 2013cs was calculate using the SiFTO light curve fitting code \citep{2008ApJ...681..482C} to produce a stretch value and the \textit{B -- V} colour at maximum. No early-time photometric measurements are available for the five remaining objects in our sample.

\subsection{Literature sample}
\label{lit_deets}
To perform a comparison with literature spectra, we have compiled a sample of late-time spectra with phases that match those of our XShooter data ($\sim$200--430 d). The late-time spectra come from literature spectral compilations \citep{2010Natur.466...82M,Blo12,2013MNRAS.430.1030S,2015MNRAS.454.3816C}, along with additional spectra of SN 2010ev \citep{2016A&A...590A...5G} and SN 2011fe \citep{2015MNRAS.450.2631M,2015MNRAS.454.1948G}.  \SiII\ 6355 \AA\ velocities within $\pm$5 d of  maximum were also complied from the literature. The majority of the \SiII\ velocities are from spectra from \cite{Blo12}, supplemented with values for individual objects: SN 2003gs \citep{suniauc8171}, SN 2010ev \citep{2016A&A...590A...5G}, SN 2011by \citep{2013MNRAS.430.1030S}, SN 2011fe \citep{2014MNRAS.444.3258M}, SN 2013dy \citep{2015MNRAS.452.4307P}, and SN 2014J \citep{2014MNRAS.445.4427A}.

\begin{figure*}
\includegraphics[width=16cm]{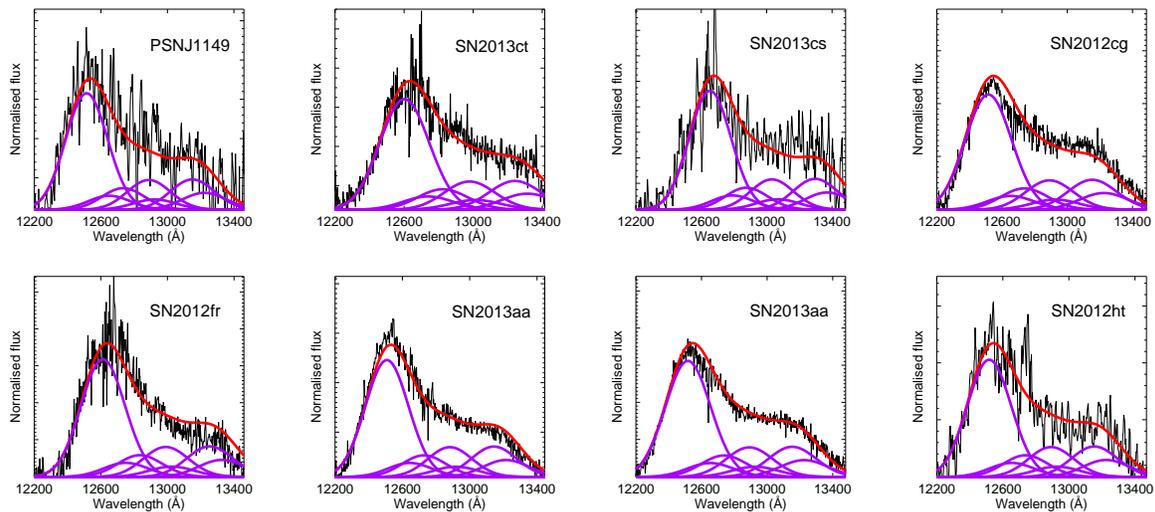} 
\caption{Best fits to the \FeiiF\ 12600 \AA\ region for the XShooter sample. The observed spectra are shown in black, the overall fits are shown in red, and the individual  \FeiiF\ features are shown in purple.  }
\label{fe_126m}
\end{figure*}

\section{Line-fitting analysis of late-time spectra}
\label{sec:line_fitting}
The late-time optical and NIR spectra of SNe Ia are dominated by emission lines of Fe-group elements, \FeiiiF, \FeiiF, \CoiiiF, and \NiIIF\ lines \cite[e.g.][]{1980PhDT.........1A,1980ApJ...239..257M}. Our aim was to use a line-fitting code on individual Fe-group spectral regions to determine the contribution of different elements to the main emission line groups seen in the optical and NIR spectra of a sample of SNe Ia. We were interested in three regions of their optical and NIR spectra: i) the 12600 \AA\ region that is dominated by \FeiiF\ lines, ii) the 7300 \AA\ region that has contributions from \FeiiF\ and \NiIIF, as well as a potential contribution from \CaiiF, and iii) the 6000 \AA\ region, which is dominated by \CoiiiF\ emission. The lines identified to be important in the 7300 \AA\ region are the same as for core-collapse SN case studied in \cite{2015MNRAS.448.2482J}, and we follow a similar analysis procedure, detailed below.

For \FeiiF\ and \NiIIF, the line-fitting code uses the latest atomic data (e.g.~rest wavelengths, transition probabilities) from the \textsc{cmfgen} compilation \citep{1998ApJ...496..407H,2012MNRAS.424..252H}, which includes the forbidden line data from \cite{1996A&AS..119...99Q} and \cite{1996A&AS..120..361Q}. For  \CoiiiF\ we use the atomic data from \cite{2016MNRAS.462.3350T}. 

Using the Sobolev formula and reasonable numbers for the densities and velocity gradients, we found that an optically-thin approximation for the relative line strengths is appropriate in this wavelength range. Assuming local thermodynamic equilibrium (LTE), the relative line strengths were calculated from temperatures in the range of 3000--8000 K, which is the expected range for SNe Ia at these phases  \citep{2015ApJ...814L...2F}. The values used in the fits are the mean values of the line strengths in this temperature range.   Line strengths are calculated relative to the strongest feature in a wavelength region.  Lines that have a strength of $<$5 per cent of the strongest line in the region are excluded since they have a negligible influence on the fit. For the three main spectral regions discussed, the strongest two lines in each case come from the same upper energy level and therefore, their relative strengths are temperature independent.

\subsection{Fitting method}
Gaussian profiles are used to fit the emission lines of the Fe-group elements since the emitting regions of these features are expected to be centrally concentrated.  The fitting of the emission features was performed in the following way. Continuum points to the red and blue of the spectral region under consideration were chosen interactively. These are difficult to set automatically because the velocity offsets are found to vary between objects, as well as due to changes in the background levels depending on the epoch of the spectrum. The relative strengths of lines of the same element and ionisation state were constrained using the radiative transition rates and the assumption of LTE level populations, as detailed above. The widths and velocities (relative to the rest wavelength) of features of the same element and ionisation state in the same spectral region were set to the same value since they are expected to be produced by the same regions. Loose priors were placed on some values as described below for the individual spectral regions.

The uncertainties in the fitting regimes were calculated using a Monte-Carlo approach. The largest uncertainties arise from the placement of the continuum points to the red and blue of the features under investigation. To produce an estimate of the uncertainties from this, the interactively chosen continuum values were varied by random amounts up to 10 \AA\ to the blue and red of this value and the fitting was then repeated. This was performed 3000 times to produce mean fit values and relative uncertainties. An additional uncertainty on the velocity measurements of 200 \kms\ was added to account for peculiar velocity effects of the host galaxies. The best fit values for the spectral regions studied, in terms of the velocity offset, full width at half maximum (FWHM), and pseudo-equivalent width (pEW),  are presented in Table \ref{tab:bestfit}.

The $\sim$4000 -- 5700 \AA\ region shows the strongest Fe emission features. However, it is also the most complex with contributions from both \FeiiF\ and \FeiiiF\ lines that are difficult to disentangle. The ratio of the strongest \FeiiiF\ lines (4658, 5270, 5011 \AA) are also sensitive to temperature changes.   Previous detailed spectral modelling of this region has failed to accurately model it, despite reasonably good agreement at other wavelengths \citep{2011MNRAS.418.1517M,2015MNRAS.450.2631M,2015ApJ...814L...2F}.  We attempted to fit this region but the fits were in general not good. Therefore, we leave further analysis of this region to the detailed simulations of Shingles et al.~(in prep.). For comparison with other works \cite[e.g.][]{2016MNRAS.462..649B}, we used the position of the peak as a measure of the velocity of the 4700 \AA\ feature (using a fit with just two lines with rest wavelengths of 4658 and 4702 \AA). 

We also excluded the $\sim$15000--18000 \AA\ region from detailed analysis due to the presence of \CoiiiF\ and \CoiiF\ lines in this region, as well as because the strongest \FeiiF\ features in this region do come from the same upper energy levels (unlike the case in the other regions studied). This wavelength region is the focus of Fl\"ors et al.~(in prep.).

\begin{figure*}
\includegraphics[width=17cm]{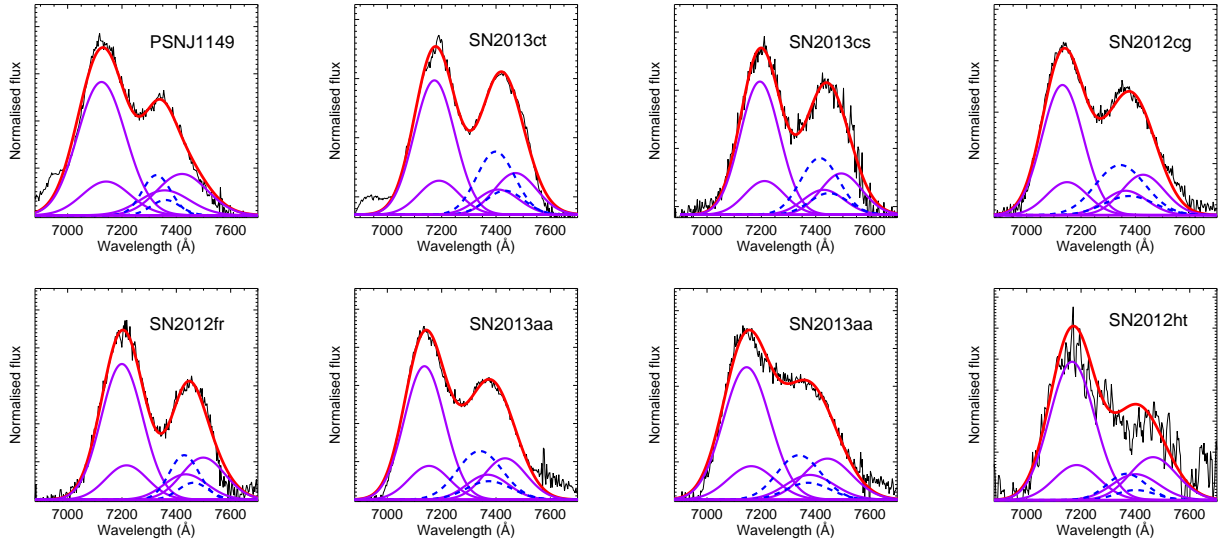} 
\caption{Best fits to the 7300 \AA\ region containing \FeiiF\ and \NiIIF\ features. The observed spectra are shown in black, the overall fits are shown in red, the  \FeiiF\ features are shown as solid purple lines, and the \NiIIF\ features are shown as blue dashed lines.}
\label{fig:feni}
\end{figure*}

\begin{figure*}
\includegraphics[width=17cm]{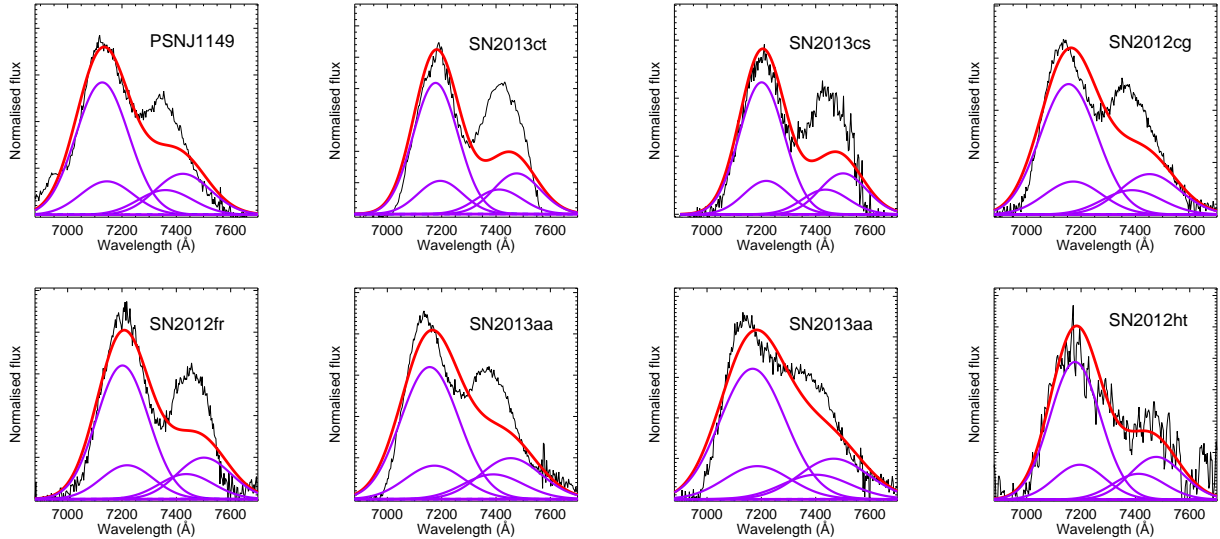} 
\caption{Best fits to the 7300 \AA\ region with no \NiIIF\ included in the fit (just \FeiiF). The observed spectra are shown in black, the overall fits are shown in red, and the \FeiiF\ features are shown in purple.}
\label{fig:feni_noni}
\end{figure*}

\begin{table*}
  \caption{Best fit values for the line-fitting analysis. }
 \label{tab:bestfit}
\begin{tabular}{@{}lccccccccccccccccccccccccccccc}
  \hline
  \hline
Name&Phase&Line & Velocity shift & FWHM & pEW &Name&Phase&Line & Velocity shift & FWHM & pEW\\
&(d)&(\AA)&(\kms)&(\kms) & (\AA) &&(d)&(\AA)&(\kms)&(\kms)&(\AA) \\
\hline
PSNJ1149&$+$206&\FeiiF\ 12567  &$-$1272$\pm$478  &7112$\pm$475&2383$\pm$780&SN2012fr&$+$357&\FeiiF\ 12567&1091$\pm$691 &7403$\pm$667 &9859$\pm$3017\\
&&\FeiiF\ 7155 & $-$1271$\pm$267 &8748$\pm$332& 618$\pm$61&  &&\FeiiF\ 7155 & 1865$\pm$205& 7498$\pm$80 &  661$\pm$28\\
&&\NiIIF\ 7378 & $-$2074$\pm$437 &5063$\pm$926&157$\pm$23 & &&\NiIIF\ 7378&2063$\pm$222 &6163$\pm$46 &  217$\pm$14\\
&&\CoiiiF\ 5890 & $-$944$\pm$201  &9346$\pm$42&1135$\pm$31&&&\CoiiiF\ 5890  &103$\pm$267&11405$\pm$265 &  360$\pm$ 32\\
SN2013ct&$+$229&\FeiiF\ 12567  &834$\pm$267  &8020$\pm$164&1941$\pm$98&SN2013aa&$+$360&\FeiiF\ 12567&$-$1408$\pm$323&7211$\pm$162 & 6234$\pm$1601 \\
&&\FeiiF\ 7155& 765$\pm$200 &7343$\pm$ 55&417$\pm$7 &  &&\FeiiF\ 7155 &$-$771$\pm$206& 7349$\pm$105 &  642$\pm$33 \\
&&\NiIIF\ 7378& 845$\pm$206 &6674$\pm$25&256$\pm$5 & &&\NiIIF\ 7378&$-$1535$\pm$266&7933$\pm$25  & 366$\pm$10 \\
&&\CoiiiF\ 5890 & 35$\pm$216 &8545$\pm$175&789$\pm$41&&&\CoiiiF\ 5890  &$-$222$\pm$258&9671$\pm$336  & 250$\pm$23\\
SN2013cs&$+$303&\FeiiF\ 12567  & 2060$\pm$770 &7135$\pm$213&2798$\pm$654& SN2013aa&$+$425&\FeiiF\ 12567 &$-$1189$\pm$637&7781$\pm$ 900&9763$\pm$3504 \\
&&\FeiiF\ 7155&  1656$\pm$207 &7122$\pm$50&406$\pm$5 &  &&\FeiiF\ 7155 & $-$385$\pm$244&8495$\pm$210 &  738$\pm$46  \\
&&\NiIIF\ 7378&  1496$\pm$207 &6765$\pm$25 &236$\pm$2 & &&\NiIIF\ 7378&$-$1613$\pm$395& 7718$\pm$594 &  329$\pm$25\\
&&\CoiiiF\ 5890 & $-$254$\pm$216 &11098$\pm$328&1493$\pm$271&&&\CoiiiF\ 5890  &--&-- &-- &\\
SN2012cg&$+$339&\FeiiF\ 12567  & $-$1203$\pm$422 &8001$\pm$1857&6770$\pm$1211&SN2012ht&$+$433&\FeiiF\ 12567&$-$1179$\pm$469&7497$\pm$1270 & 3830$\pm$1444 \\
&&\FeiiF\ 7155& $-$981$\pm$201 &7474$\pm$47 &  474$\pm$11 &  &&\FeiiF\ 7155 & 494$\pm$381& 8130$\pm$693 &  893$\pm$161 \\
&&\NiIIF\ 7378&  $-$1381$\pm$212&8801$\pm$30&309$\pm$3 & &&\NiIIF\ 7378&424$\pm$2628 &6783$\pm$2826&174$\pm$34&\\
&&\CoiiiF\ 5890  &$-$65$\pm$256  &11906$\pm$192  & 264$\pm$16&&&\CoiiiF\ 5890  &--&-- &-- &\\
\hline
\end{tabular}
 \begin{flushleft}
 \end{flushleft}
\end{table*}

\subsection{12600 \AA\ \FeiiF-dominated region fitting}
\label{12600_fitting}
The 12600 \AA\ region is dominated by the \FeiiF\ 12567 \AA\ feature, along with six other weaker \FeiiF\ lines (12704, 12788, 12943, 12978, 13206, 13278 \AA) that have $<$30 per cent the strength of the 12567 \AA\ line. No lines of other elements or other ionisation states of Fe are predicted to be present in this region, making it an excellent choice for constraining the properties of the \FeiiF\ emitting region. The second strongest line at 13206 \AA\ comes from the same energy level as the strongest line so their ratio is temperature independent. Assuming LTE, temperature-induced changes in the relative line strength of the five weakest lines are of the order of $\sim$5 per cent, resulting in variations in the strength of the strongest 12567 \AA\ lines of $\sim$5 per cent, less than the uncertainty on the fit.

 As we have constrained the relative strengths from the atomic data, and the velocity shifts and widths of the \FeiiF\ features are tied, there are just four free parameters for the fit to the region (strength, width, velocity offset, and the height of the continuum). No priors are placed on these values. The best fits to this 12600 \AA\ \FeiiF-dominated region are shown in Fig.~\ref{fe_126m}.

\subsection{7300 \AA\ \FeiiF/\NiIIF-dominated region}
\label{feni_fitting}
The 7300 \AA\ line complex in late-time SN Ia spectra is made up of lines of \FeiiF\ (7155, 7172, 7388, 7453 \AA), \NiIIF\ (7378, 7412 \AA), and possibly \CaiiF\ (7291, 7324 \AA).  Different explosion mechanisms predict different contributions of these elements, depending on the presence of stable Fe-group material and the completeness of burning in the explosion. 

The relative strengths of the two strongest lines of the \FeiiF\ lines are temperature independent. There is a small temperature dependence of the two weaker lines relative to the strongest line, \FeiiF\ 7155 \AA, but changes in temperature in the range 3000--8000 K results in a change in the relative strength of the strongest line of $<$5 per cent. These \FeiiF\ lines are expected to be dominated by emission of the decay of radioactive $^{56}$Co. A smaller contribution from stable primordial Fe ($^{54}$Fe, $^{56}$Fe) may be present but it is impossible to disentangle these contributions. The initial guess for the velocity and width of \FeiiF\ lines in the 7300 \AA\ complex were set as the output of the \FeiiF\ 12600 \AA\ region but were allowed to vary within $\pm$5000 \kms\ of the NIR \FeiiF\ velocity and width values (much greater than the velocity shifts or the range of widths observed).

Since $^{56}$Ni decays with a half-life of 6.1 days, the contribution from radioactive Ni at late-times is negligible. Therefore, any Ni emission lines seen in late-time spectra must come from stable $^{58}$Ni (and to a much smaller extent from stable $^{60}$Ni). For example, the mass ratio of  $^{60}$Ni to $^{58}$Ni is $\lesssim$10 per cent for the  M$_{ch}$ delayed-detonation (DDT) models of \cite{2013MNRAS.429.1156S} so we refer to $^{58}$Ni to mean the total stable Ni contribution.  Two \NiIIF\ features (7378, 7412 \AA) are expected to emit in this region if sufficient stable Ni is present. No priors were placed on the velocity offset of the \NiIIF\ lines.  Different explosion models predict different extents of the stable Fe-group element emitting regions. W7 predicts that the width of the stable Fe-group element emitting region should be $<$6000 \kms\ \citep{1999ApJS..125..439I}, while sub-M$_{ch}$ models prefer values of $<$7000 \kms \citep{2010ApJ...714L..52S}. To be agnostic as to the explosion mechanism, the width of the \NiIIF\ was left to vary between 2000 \kms\ and the width of the radioactive-chain dominated \FeiiF\ 12567 \AA\ line ($+$10 per cent), which is typically of the order of 7000--9000 \kms.  

Fig. \ref{fig:feni} shows the best fit to the 7300 \AA\ \FeiiF/\NiIIF\ region under the assumptions and constraints detailed above. Since constraining the contribution of \NiIIF\ to the region was one of our primary aims, we repeated the fitting removing the two \NiIIF\ features from the fit and obtained the results shown in Fig.~\ref{fig:feni_noni}. It can be seen that the fits are significantly worse for all SNe apart from SN 2012ht (this is because it contained only a small contribution of \NiIIF\ in the best fit when the presence of \NiIIF\ was allowed and is also the lowest S/N of our spectral sample). Therefore, for 6 out of 7 SNe in our sample, a non-negligible contribution from \NiIIF\ emission is required to produce a reasonable fit to the region.

\CaiiF\ features (7291, 7324 \AA) could also be present in this wavelength region.  The Ca-emitting regions are expected to be further out in the ejecta than the Fe-group elements with a shell-like structure for most explosion scenarios for normal luminosity SNe Ia that are studied in this work. We approximated this as a shell with a width in the range, 5000 -- 15000 \kms\ covering the range predicted by Chandrasekhar and sub-Chandrasekhar mass explosions. We used a boxy profile with tapered edges with the flux for a given wavelength of the form, F$_{\lambda} =$ $1-$ $\Delta v^{2}$, where $\Delta v$ is the velocity shift from the rest velocity \citep{Jerkstrand2017}. We tested the contribution of the \CaiiF\ lines to the fit by performing the fits again with a contribution from \CaiiF. To test if the fit is strengthened by including \CaiiF, we used the Bayesian Information Criterion \citep[BIC;][]{schwarz1978} to select between models with and without a contribution from \CaiiF. The BIC penalises models with a greater number of parameters. The model with the lower information criterion score is considered to be the better model for explaining a given dataset. We measured the BIC for models with and without a contribution from \Caii\ and find $\Delta$BIC values of $\gg$10 in favour of the no Ca model, providing strong evidence against the need to include \CaiiF\ in the fit to this region. 

 \begin{figure}
\includegraphics[width=8.8cm]{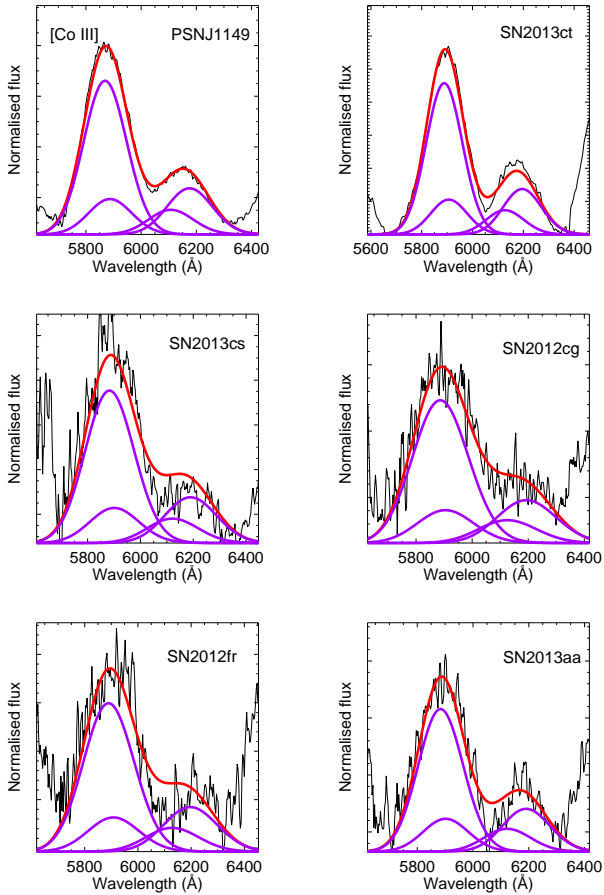} 
\caption{Best fits to the 6000 \AA\ region for the four \CoiiiF\ lines. The observed spectra are shown in black, the overall fits are shown in red and the four individual \CoiiiF\ profiles are shown in purple.}
\label{fig:coiii}
\end{figure}

If we assume instead that \CaiiF\ emission was responsible for the `\NiIIF-feature', the redder peak in this region, and that it was allowed to have a Gaussian instead of boxy profile (although this is not expected for the main explosion models), then the velocity offsets to fit the `\NiIIF-feature' would be shifted to the red by a further $\sim$3500 \kms. For the most extreme redshifted case of SN 2012fr, this feature would have a velocity offset from the rest wavelength of \CaiiF\ of $\sim$5500 \kms. When the red peak is assumed to be dominated by \NiIIF, then the distribution of velocity offsets is centred on zero with three of the XShooter sample offsets lying to the blue and five to the red. If the origin of this feature was \CaiiF\ emission then all the velocity shifts of the sample would be shifted strongly towards the red, again arguing against a centrally-peaked Ca distribution being the origin of this feature.

\subsection{6000 \AA\ \CoiiiF-dominated region}
\label{coiii_fitting}
The $\sim$5700 -- 6300 \AA\ region displays two peaks dominated by \CoiiiF\ emission that radioactively decays with time from $\sim$60 d after explosion \citep{2015MNRAS.454.3816C}, and are undetectable after $\sim$400 d (Fig.~\ref{fig:coiii}). Since there are no non-negligible stable Co isotopes, the \CoiiiF\ lines trace regions rich in radioactive material. Along with composition changes, the regions emitting \CoiiiF\ compared to \FeiiF\ and \NiIIF\ lines are also likely to differ in their thermodynamic conditions since they result in the production of different ionisation states. The strengths of the \CoiiiF\ lines (5890, 5908, 6129, 6197 \AA) are constrained relative to the strongest feature. The second strongest line at 6197 \AA\ has a fixed ratio to the strongest feature because it comes from the same upper energy level. The weakest two lines have a small temperature dependence in the 3000--8000 K regime, causing a $\sim$6 per cent uncertainty in the relative strength of the strongest line. No constraints were placed on the strength of the strongest line, its velocity offset, or width. In agreement with \cite{2014MNRAS.439.3114D}, we find that this spectral region is well fit without a contribution from \Nai\ D.

\begin{figure}
\includegraphics[width=8.4cm]{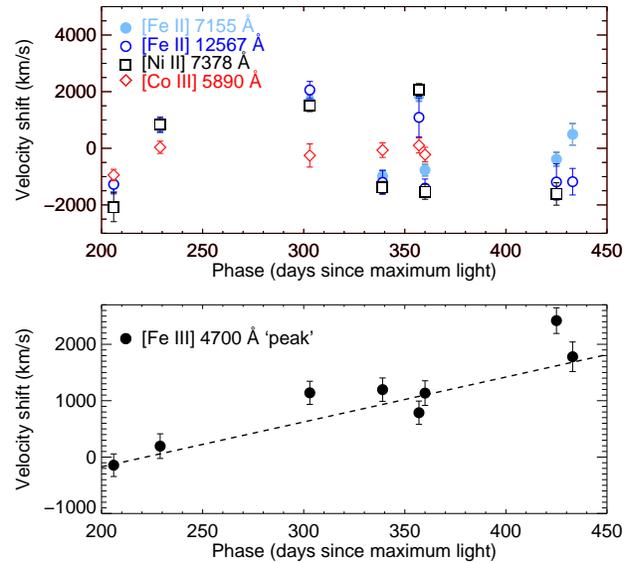} 
\caption{Velocity shift with respect to their rest wavelength versus phase for the \FeiiF\ 7155 \AA\ (pale blue solid circles), \FeiiF\ 12567 \AA\ (blue open circles), \NiIIF\ 7378 \AA\ (black squares), \CoiiiF\ 5890 \AA\ (red squares) in the top panel, and \FeiiiF\ 4700 \AA\ features in the bottom panel. The \FeiiF\ 7155 \AA, \FeiiF\ 12567 \AA, and \CoiiiF\ 5890 \AA\ velocities are from full fitting to the region, while the \FeiiiF\ 4700~\AA\ values are from a `peak' fit to the strongest emission feature in this region. The best-fit line to the \FeiiiF\ is shown, demonstrating a shift redward with time of this feature, as has been previously identified. }
\label{fig:vel_shift_phase}
\end{figure}

\begin{figure*}
\includegraphics[width=15cm]{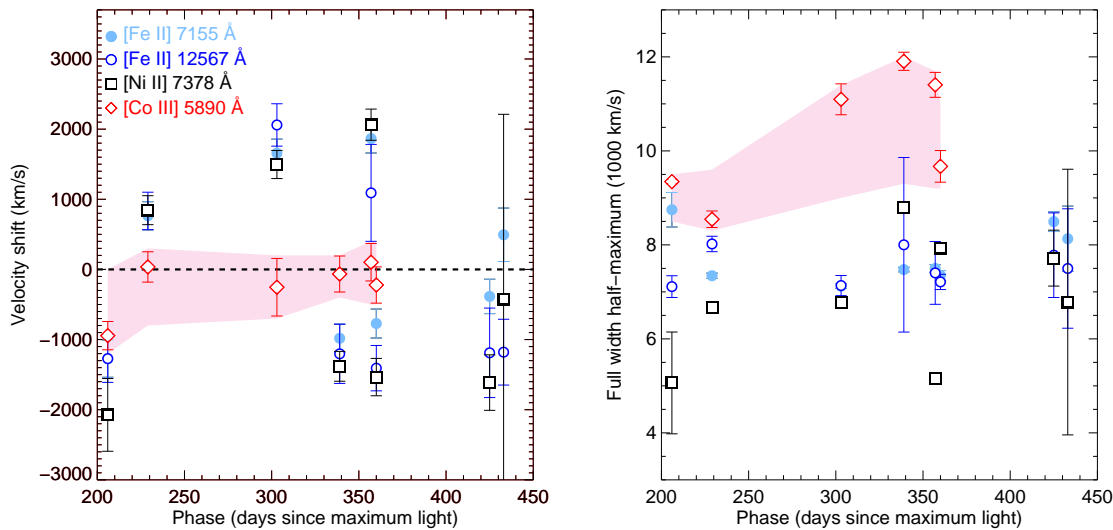} 	
\caption{Left: Velocity shifts of the \FeiiF\ 7155 \AA, \FeiiF\ 12567 \AA\, \CoiiiF\ 5890 \AA, and \Nii\ 7378 \AA\ features as a function of time. The pink shaded region highlights the smaller relative velocities for the \CoiiiF\ lines. Right: FWHM of the \FeiiF\ 7155 \AA, \FeiiF\ 12567 \AA\, \CoiiiF\ 5890 \AA, and \Nii\ 7378 \AA\ features as a function of time. The pink shaded region highlights the larger FWHM values for the \CoiiiF\ lines.  }
\label{fig:vel_width_shift}
\end{figure*}

\section{Late-time velocity shifts and line widths}
\label{sec:shifts}
An asymmetric distribution of Fe in SNe Ia was first suggested by \cite{2004A&A...426..547S} by studying the shapes of NIR \FeiiF\ lines in late-time spectra. Velocity shifts of the central wavelength of the \FeiiF\ 16440 \AA\ feature with respect to its rest wavelength were identified by \cite{2006ApJ...652L.101M} for three SNe Ia, 2003du, 2003hv and 2005W. In SN 2003hv, a shift was observed for \CoiiiF\ at mid-infrared wavelengths \citep{2007ApJ...661..995G} and \FeiiF\ at optical wavelengths \citep{2009A&A...505..265L}.  Optical line shifts have also been identified for the \FeiiF\ 7155 \AA\ and \NiIIF\ 7378 \AA\ features with peak velocity shifts from rest of $\pm$3000 \kms\  that correlate with \SiII\ velocities measured at maximum light \citep{2010Natur.466...82M,2010ApJ...708.1703M,Blo12,2013MNRAS.430.1030S}.  Here we discuss the velocity shifts, as well as measurements of the line widths, for our sample of late-time spectra, and also perform a comparison with literature objects.

\subsection{Velocity shift evolution with phase}
\label{sec:shifts_phase}
In the top panel of Fig.~\ref{fig:vel_shift_phase}, the velocity shifts of the 7155, 12567 \AA\ \FeiiF\ lines, as well as those of \NiIIF\ 7378 \AA\ and \CoiiiF\ 5890 \AA, are shown as a function of phase for the 8 spectra of the 7 SNe Ia in our sample. No evolution with time is identified for our sample and the \FeiiF\ and \NiIIF\ features show large positive and negative offsets that do not appear phase dependent. \cite{2016MNRAS.462..649B} investigated the velocity offset with time for the \CoiiiF\ 5890 \AA\ region for a sample of SNe Ia with multiple spectra per object and found a flat trend (within the uncertainties) of their velocities in the range 200 -- 350 d post maximum. In the bottom panel of the same plot, the peak of the \FeiiiF\ 4700 \AA\ is shown as a function of phase. This peak value was not measured from a full fit to the region but from a simple Gaussian fit to the strongest 4700~\AA\ peak. A clear trend of increasing redshift with time is identified for the \FeiiiF\ 4700 \AA\ feature. This is in agreement with previous studies, where the wavelength of the peak of the feature was found to shift towards the red with time from $\sim$50 to 450 d past maximum light \citep{2010ApJ...708.1703M,2016MNRAS.462..649B,2017MNRAS.472.3437G}. This could be due to contributions from other features in this region, such as permitted Fe and \CoiiF\ lines.  A redshifting with time of line peaks has previously been observed in a number of core-collapse SNe \citep[e.g.][]{2009MNRAS.397..677T}, and has been suggested to be caused by a decrease in opacity with time, that mainly effects bluer wavelengths \citep{2015A&A...573A..12J}.

\subsection{Velocity offsets and widths of Fe-group elements}
\label{sec:vel_width}
We wished to investigate in our  sample the relative velocity shifts of different Fe-group elements (and ionisation states) to determine if their emitting regions are the same. Previous studies have suggested that \NiIIF, \FeiiF, and \FeiiiF\ features are produced in different regions of the ejecta that depend on the density \citep{2010ApJ...708.1703M}. As shown in Section \ref{sec:shifts_phase}, the \FeiiiF\ 4700 \AA\ feature is not suitable for this study because its central position evolves with time. 

In Fig.~\ref{fig:vel_width_shift}, we show the results of our comparative study of the velocity and widths of singly and doubly ionised elements.  We have compared the velocities of the \CoiiiF\ 5890 \AA\ line to those of the \NiIIF\ 7378 \AA, \FeiiF\ 7155, 12567 \AA\ features. We find that the mean absolute offset from zero velocity ($\langle \vert v \vert \rangle$) for \CoiiiF\ is much smaller than that for the lower ionisation lines of \FeiiF\ and \NiIIF.  The \FeiiF\ 7155 and 12567 \AA\ features have mean absolute velocity offsets from zero of 1024$\pm$533 and 1279$\pm$355 \kms, respectively. The \NiIIF\ 7378 \AA\ feature has a mean absolute velocity offset of 1429$\pm$563 \kms\ that is slightly higher that the \FeiiF\ features but consistent within the uncertainties. However, the \CoiiiF\ 5890 \AA\ feature has a significantly lower mean absolute velocity offset of 271$\pm$341 \kms, which is consistent with zero offset from its rest wavelength. 

The $\sim$15000 -- 18000 \AA\ region is also dominated by \FeiiF\ features. As discussed in Section \ref{sec:line_fitting}, we have excluded this region from our detailed analysis because of potential contributions from \CoiiF\ and \CoiiiF that makes the fitting of the \FeiiF\ features less clean. However, if we assume that the main peak in this region is dominated by emission from the \FeiiF\ 16440 \AA\ line, then it can be seen from Fig.~\ref{flat_tops} that its peak also shifted in a similar way to the peak of the 12600 \AA\ region. The peaks of the 12600 and 16400 \AA\ regions are shifted consistently to the red for SNe 2013ct, 2013cs and 2012fr. This confirms our previous results that the distribution of \FeiiF\ emission lines is shifted in SNe Ia both positively and negatively with respect to the rest wavelengths of the lines.

We have also investigated the extent of the emitting regions of these Fe-group features using their FWHM (Fig.~\ref{fig:vel_width_shift}, right panel). We found that the \CoiiiF\ lines (mean FWHM of 10300$\pm$1300 \kms) are significantly broader than the  \FeiiF\ 7155 \AA, \FeiiF\ 12567 \AA, and \NiIIF\ 7378 \AA\ lines, with mean FWHM of 7800$\pm$600, 7500$\pm$400, and 6900$\pm$1300 \kms, respectively. 

\subsection{Late-time velocity shifts of \FeiiF\ and the connection to maximum-light \SiII\ velocities}

A connection between the average peak line shift of the two main components of the 7300 \AA\ region (suggested to be due to \FeiiF\ 7155 \AA\ and \NiIIF\ 7378 \AA\ for the blue and red peaks, respectively) in the late-time spectra (150 -- 400 d past max.) and the \SiII\ velocity gradient at maximum light was first identified by \cite{2010Natur.466...82M}. SNe Ia displaying redshifted late-time 7300~\AA\ region spectral features have higher velocity gradients at early times, suggested to originate from asymmetric and off-centre explosion mechanisms. \cite{2013MNRAS.430.1030S} extended this work to show that the relation also holds if \SiII\ velocities at maximum are used instead of the harder to measure velocity gradient. 

We wished to measure velocity offsets for our new late-time spectral sample ($>$200 d after max.) and compare with \SiII\ velocities at maximum (presented in Table \ref{tab:SN_info}). We included the three additional XShooter spectra (described in Section \ref{sec:obs_data}), with lower S/N spectra in the NIR region, to increase the sample.  Unlike \cite{2010Natur.466...82M}, we chose to focus on the blueward peak that is predominantly \FeiiF\ because the redward peak of this feature is not due to solely \FeiiF\ or \NiIIF\ but is a blend of both (Fig.~\ref{fig:feni}). Therefore, it is not as clean a measurement of Fe-group velocity offsets as the \FeiiF\ 7155 \AA\ peak. 

We measured the \FeiiF\ 7155 \AA\ velocity shift using two methods: the `peak' method of \cite{2010Natur.466...82M} and the more detailed multi-line fitting method we presented in Section \ref{sec:line_fitting}. We also remeasured values of the late-time velocity offset of the \FeiiF\ 7155 \AA\ feature for the literature sample described in Section \ref{lit_deets}, which includes SNe Ia falling in the same phase range of our late-time sample of $\sim$200--430 d post max. This is a stricter lower phase cut than that used by \cite{2010Natur.466...82M}.

\begin{figure}
\includegraphics[width=9.2cm]{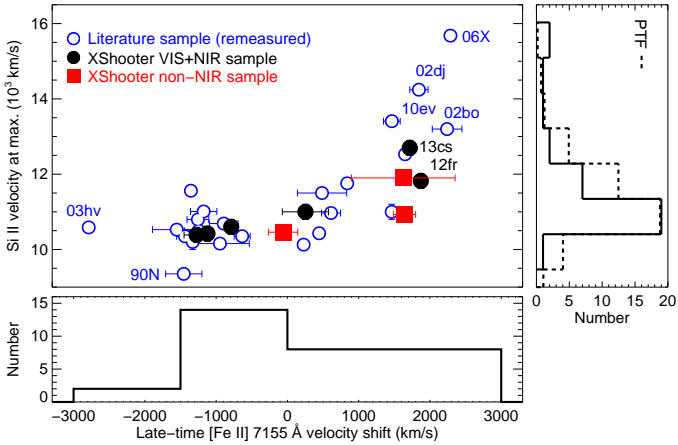} 	
\caption{Main: \SiII\ velocity at maximum light versus the velocity shift of the \FeiiF\ 7155 \AA\ feature from spectra at $>$200 d after maximum (measured from a multi-Gaussian fit to the feature). The new XShooter sample with high S/N NIR spectra are shown in black solid circles, the lower XShooter S/N sample without good NIR data is shown as red squares, and the remeasured literature sample is shown as blue open circles. Bottom: Histogram of the late-time \FeiiF\  7155 \AA\ velocity shifts for the combined XShooter and literature samples (black solid line). Side: Histogram for the \SiII\ velocity at maximum light for the combined sample (solid black line), with the PTF SN Ia sample of 247 SNe Ia within $\pm$5 d of maximum light from \protect \cite{2014MNRAS.444.3258M} normalised to the peak of the combined sample (black dashed line).}
\label{fig:vel_corr}
\end{figure}

A comparison between the two methods (`peak' and full fit) shows that there is a systematic offset such that the values measured from the peak of the blueward feature (\FeiiF\ 7155 \AA) are $\sim$150--300 \kms\ offset to the red compared to the full fitting method. This is evidence for line blending skewing the measured peak value of the emission profile. However, this is a small offset and generally within the uncertainties of the line fitting so we conclude that the choice of method does not significantly impact the final results.  We used the full fitting method in our further analysis.

Fig.~\ref{fig:vel_corr} shows the results of our analysis of the relation between \SiII\ velocity at maximum and late-time velocity offset of the \FeiiF\ 7155 \AA\ lines for our SN sample and remeasured literature sample. We find results in agreement with that of previous studies but strengthened (with no clear outliers) by the inclusion of additional objects, the stricter phase cut, and the consistent measurements of solely the \FeiiF\ 7155 \AA\ feature using our more detailed fitting technique. We also tested for relations between the maximum-light light curve width parameter, `stretch', the \textit{B-V} colour at maximum, and the late-time velocities but find no significant correlations.

A similar relation is seen between \SiII\ velocity at maximum and the late-time velocity offset of the NIR \FeiiF\ 12567 \AA\ line -- not surprisingly given the clear correlation between the optical and NIR \FeiiF\ lines observed in Fig.~\ref{fig:vel_width_shift}. However, the sample size is small since the literature spectral sample generally did not cover combined optical and NIR wavelengths. We have also investigated the relation between the \SiII\ velocity at maximum and the late-time velocity shift of the \CoiiiF\ 5890 \AA\ lines for the combined XShooter and literature sample. We identify a tentative trend in the sense that SNe Ia with positive (redward shifted) \CoiiiF\ late-time velocities are more likely to have higher  \SiII\ velocities at maximum (in the same sense as the \FeiiF\ lines). However, when we measure the mean \SiII\ velocity for samples with redward or blueward \CoiiiF\ shifts, the means agree within the uncertainties. The \CoiiiF\ lines also have significantly smaller offsets from zero velocity (Fig.~\ref{fig:vel_width_shift}).

To determine if the combined XShooter and literature sample is unbiased with respect to selection effects, we have compared it to the untargeted Palomar Transient Factory (PTF) sample of 247 SNe Ia within 5 d of maximum light \citep{2014MNRAS.444.3258M}. We have found that the percentage of SNe Ia with \SiII\ velocities above 12000 \kms\ is consistent between the samples at $\sim$18 and $\sim$16 per cent for the late-time spectral sample and the maximum-light PTF sample, respectively. This suggests that there is no significant bias in the sample selection for the  late-time sample.

\section{Late-time colours of SNe Ia}
\label{sec:colour}
\begin{figure}
\includegraphics[width=8.5cm]{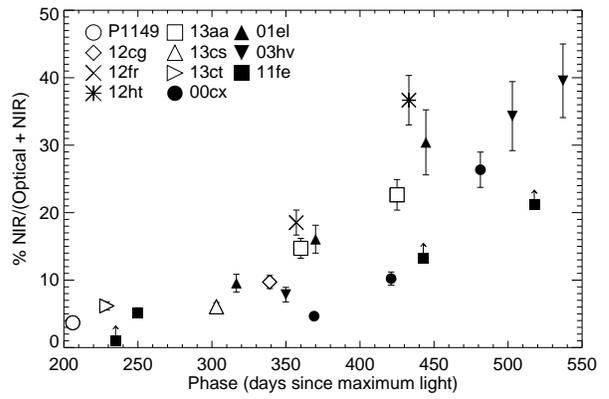} 
\caption{Percentage flux through the  NIR \textit{JHK} filters compared to the combined flux from a broad optical filter  (3550 -- 9500 \AA)  and the \textit{JHK} filters. Three epochs of data for SN 2011fe are marked as lower limits because only \textit{J} band data are available as the NIR contribution. }
\label{nir_percent}
\end{figure}

The percentage flux in the NIR bands was compared to the total flux in the optical and NIR by measuring the flux through a broad optical filter (3550--9500 \AA), as used in \cite{2017MNRAS.468.3798D}, and the sum of the flux through the \textit{JHK} bands. The percentage flux in the NIR compared to the combined optical+NIR is shown as a function of phase in the range of 200 -- 550 d post maximum in Fig.~\ref{nir_percent}. This figure includes measurements from the new spectral sample presented here, as well as the late-time optical and NIR measurements of SNe 2000cx, 2001el, 2003hv, 2011fe \citep{2004A&A...428..555S,2007A&A...470L...1S,2009A&A...505..265L} that were presented in \cite{2017MNRAS.468.3798D}. 

An increase in the NIR flux contribution with time is seen over the phase range studied. The NIR contribution to the combined optical and NIR flux is relatively constant in the range of 200 -- 350 d with a mean of 7$\pm$2 per cent. In the range 350 -- 450 d, there is a larger spread with a mean of 13$\pm$10 per cent, while in the range 450 -- 550 d, the NIR flux contribution has increased significantly with a mean of 33$\pm$7 per cent.

\section{Calculation of Ni to Fe abundance ratio}
\label{sec:nife_calc}

\begin{figure*}
\includegraphics[width=14cm]{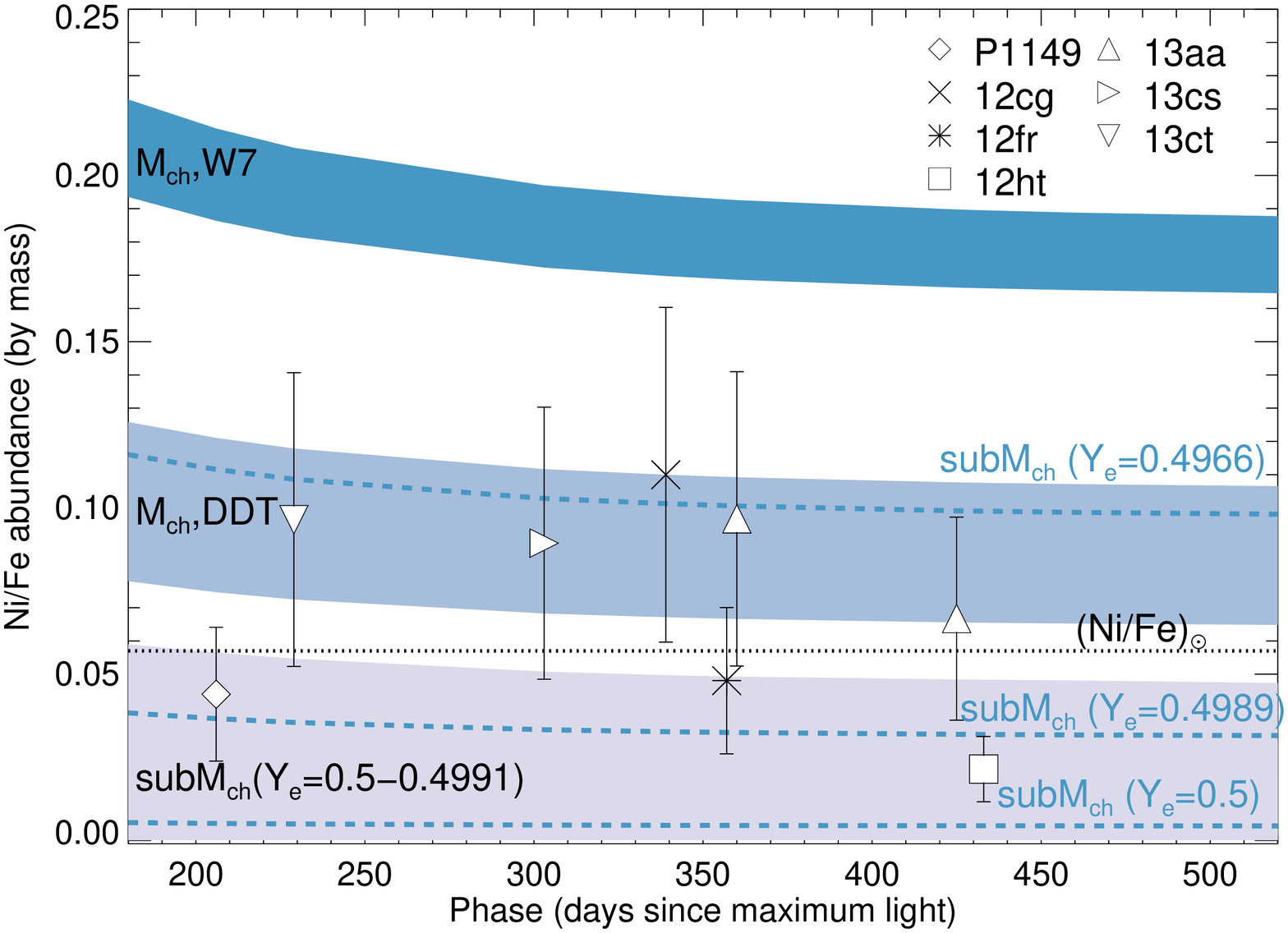} 
\caption{Ratio of Ni to Fe abundance (by mass) compared to bands of the M$_{ch}$ `DDT' models \protect \citep{2013MNRAS.429.1156S}, and `W7' models \protect \citep{1997NuPhA.621..467N,1999ApJS..125..439I}. The sub-M$_{ch}$ detonation models of \protect \cite{2010ApJ...714L..52S} and \protect \cite{2013MNRAS.429.1425R} are marked as blue dashed lines (subM$_{ch}$) at Y$_e$ = 0.5, 0.4989, and 0.4966 (equivalent to $\sim$0, 2 and 6 times solar metallicity, respectively). The sub-M$_{ch}$ detonation models of \protect \cite{2018ApJ...854...52S} in the range of Y$_e$ = 0.5 -- 0.4991  (equivalent to $\sim$0--2 times solar metallicity) are marked as a solid band (subM$_{ch}$). For reference, the solar metallicity from \protect \cite{2009ARA&A..47..481A} is equivalent to Y$_e$=0.499 and the solar Ni/Fe abundance ratio is 0.057 (black dotted line). }
\label{nife_abund}
\end{figure*}

We have produced fits to the 7300 \AA\ spectral region, which output relative line strengths, velocity shifts and line widths of \FeiiF\ and \NiIIF. The pseudo-equivalent widths of the line profiles can be used to estimate the Ni/Fe abundance for SNe Ia in the sample, under some basic assumptions. Assuming optically thin emission, we used equation 2 of \cite{2015MNRAS.448.2482J}, with an additional term to account for the departure coefficients, to give the equation

\begin{equation}
\frac{L_{7378}}{L_{7155}} = 4.9\ \frac{n_{\NiII}}{n_{\Feii}} \exp  \bigg(  \frac{0.28}{kT} \bigg) \ \frac{dc_{\NiII}}{dc_{\Feii}}
\end{equation}
where $L_{7378}/{L_{7155}}$ is the measured flux ratio of the \NiIIF\ 7378 \AA\ to \FeiiF\ 7155 \AA\ line, $n_{\NiII}$ and $n_{\Feii}$ are the number densities of \NiII\ and \Feii, $k$ is the Boltzmann constant in units of eV per K, T is the temperature in K, and $dc_{\NiII}/dc_{\Feiii}$ is the ratio of departure coefficients from LTE for the appropriate states of  \NiII\ and \Feii, respectively.  The numerical constants are the same as were used in \cite{2015MNRAS.448.2482J}.

The ratio of departure coefficients for the Fe and Ni-emitting zones is estimated from the W7 model at 330 d using the modelling of  \cite{2015ApJ...814L...2F} to be in the range of 1.2 -- 2.4 over the phase range of interest with smaller coefficients at earlier times when the conditions are slightly closer to LTE (priv.~comm.). We use a mean value of the departure coefficient of 1.8. Since \NiII\ and \Feii\ are both singly ionised and have similar ionisation potentials, we assumed that the $n_{\NiII}$/$n_{\Feii}$ ratio should closely follow the total Ni/Fe ratio. This is supported by the nebular phase modelling of \cite{2015ApJ...814L...2F} at 330 d past maximum light, which gives values in the range of 0.8 -- 1.2, with a mean of 1.0 (priv.~comm.). Similar ionisation rates are also found in the NLTE modelling of Shingles et al. (in prep.).

The uncertainties on the Ni/Fe ratio were calculated by varying the temperature in the range 3000-8000 K and the departure coefficient in the range 1.2 -- 2.4  \cite[the range of departure coefficients in different zones from the model of][]{2015ApJ...814L...2F}, and the ionisation balance in the range of 0.8 -- 1.2, also from the \cite{2015ApJ...814L...2F} model. The uncertainties on the line fitting are negligible in comparison to these uncertainties.
 
Fig.~\ref{nife_abund} shows the measured ratio of Ni to Fe abundance compared to four SN Ia explosion models (a M$_{ch}$ deflagration model, a M$_{ch}$ DDT model, and two different versions of sub-M$_{ch}$ detonation models).  The M$_{ch}$ deflagration explosion model, `W7', is shown with a range of values from \cite{1997NuPhA.621..467N} and \cite{1999ApJS..125..439I}, along with the three-dimensional M$_{ch}$ DDT models, `N100', `N100l', `N100h', `N150', of \cite{2013MNRAS.429.1156S}. In the DDT models, the neutron-rich isotopes, such as $^{58}$Ni, are synthesised mainly in the explosion itself in the highest density regions, and these models set the range of the models labelled as `DDT' in Fig.~\ref{nife_abund}.  

Neutronisation can also occur due to a increased progenitor star metallicity that results in a higher $^{22}$Ne content, which in turn introduces an excess of neutrons and different electron fractions, Y$_e$. This results in the production of neutron-rich isotopes but in smaller amounts than those produced during the explosion itself. For the M$_{ch}$ DDT models of \cite{2013MNRAS.429.1156S}, variations in initial progenitor star metallicity are dwarfed by the variations produced by varying the central density at explosion.

Two sets of sub-M$_{ch}$ detonation models are shown from \cite{2010ApJ...714L..52S} and \cite{2018ApJ...854...52S}. In these models, the neutron-rich  $^{58}$Ni depends on the progenitor star metallicity and is not produced in the explosion. The models of \cite{2010ApJ...714L..52S} are of a 1.06 \msun CO white dwarf calculated for the case of zero metallicity (Y$_e$ = 0.5), and two neutron-rich environments of Y$_e$ = 0.4989 and 0.4966, roughly equivalent to twice and six times solar metallicity, respectively. The models of \cite{2018ApJ...854...52S} are of 0.9 and 1.0 \msun\ CO white dwarfs, calculated in the range, Y$_e$=0.5 -- 0.4991, roughly equivalent to zero to twice solar metallicity. 

All the model ratios of Ni over Fe abundance decline with phase because while the stable Ni mass stays constant, the amount of Fe is increasing with time due to decay from Co. `W7' over-predicts the Ni/Fe abundance for all the SNe Ia in the sample. The data are found to be broadly consistent with the `DDT' model range, as well as some of the sub-M$_{ch}$ models. However, to explain the data, the sub-M$_{ch}$ detonation models would need neutron excesses equivalent to a few times solar metallicity.

\section{Discussion}
\label{sec:discussion}

We have presented an analysis of the largest sample to date of late-time optical and NIR SN Ia spectra, where we used a line-fitting code to estimate the widths, velocity offsets, and strengths of prominent spectral features. In this section, we review the evidence for asymmetries in the ejecta and the constraints on the presence of stable material. We also discuss the rate of cooling of the ejecta, as well as the absence of flat-topped profiles. These results are linked to a number of explosion models to constrain which models are the most viable to explain the bulk of normal SNe Ia.

\subsection{Stratification of stable and radioactive material}
\label{sec:strat}
In the W7 model (often used as a reference model), the stable material is more centrally concentrated than the radioactive decay products. The velocities of the stable material are predicted to be $\lesssim$4000 \kms\ \citep{1997NuPhA.621..467N,1999ApJS..125..439I}. The stable Ni and Fe material in SN 2011fe was found using a modified density W7 model to have velocities below $\sim$4000 \kms\ \citep{2015MNRAS.450.2631M}.   The DDT models predict distributions of stable $^{58}$Ni that when spherically averaged are approximately co-spatial with the radioactive material \citep[see fig.~1 of][]{2014MNRAS.444..350S}. The sub-M$_{ch}$ detonation models of \cite{2010ApJ...714L..52S} and \cite{2018ApJ...854...52S} predict distributions of stable Fe-group material that cover a similar velocity range to that of the radioactive isotopes.

For our SN Ia spectral sample, we have measured the width of Fe-group element features to constrain the extent of their emitting region. The width of the \CoiiiF\ line is found to be significantly broader than those of the \FeiiF\ and \NiIIF\ features. The mean FWHM of the \FeiiF\ 7155 \AA\ and \NiIIF\ 7378 \AA\ lines are 7600$\pm$700 and 6600$\pm$1500 \kms, respectively, agreeing within the uncertainties. The mean FWHM of the \CoiiiF\ 5890 line is 10300$\pm$1300 \kms. This means the region producing the  \FeiiF\ and \NiIIF\ features is smaller than the region producing the \CoiiiF\ lines.

The \NiIIF\ emission is dominated by stable $^{58}$Ni since the radioactive $^{56}$Ni is decayed by these epochs but the \CoiiiF\ lines are produced from $^{56}$Co, the product of radioactive decay. To first order, since the \NiIIF\ emitting region is narrower that of the \CoiiiF\ region, we can assume that the underlying composition between the two emitting regions is different. Since the difference in mean FWHM for the \NiIIF\ 7378 \AA\ and \CoiiiF\ 5890 \AA\ lines is of the order of a thousand to a few thousand \kms\ (taking into account the uncertainties), this suggests that the stable material is (at least slightly) more concentrated and coming from a narrower region than the radioactive material. We have not used \FeiiiF\ in our study because of the time-dependent evolution of the most prominent \FeiiiF\ feature (see Section \ref{sec:shifts_phase}). 

This suspected stratification of the stable and radioactive material is in quantitative agreement with most of the models (W7, DDT, sub-M$_{ch}$ detonation). However, W7 predicts that the velocities of the stable material are $\lesssim$4000 \kms, which we do not see in our sample. The lowest FWHM of  \NiIIF\ 7378 \AA\ for a SN Ia in our sample is SN 2012fr, with a FWHM of the \NiIIF\ 7378 \AA\ line of $\sim$4650 \kms.  The gravitationally-confined detonation models of \cite{2016A&A...592A..57S} are ruled out for the SNe Ia in our sample; these models have no stable Fe or Ni at low velocities but significant high-velocity stable material, which is at odds with our results.

However, there are a number of caveats with connecting the underlying distribution of stable and radioactive material to line emissions of \NiIIF\ and \CoiiiF. Since their ionisation states are different (singly-ionised \NiIIF\ compared to doubly-ionised \CoiiiF), the density of their emitting regions is expected to be different \citep{2010ApJ...708.1703M}.    An isolated blob of stable material will not emit without an energy source -- there must be some radioactive material mixed in with the stable material to produce line emission. Therefore, although different line emitting regions are likely dominated by differences in the composition, the role of thermodynamic conditions can of course not be neglected to focus only on differences in composition. Full non-LTE radiative-transfer modelling of different explosion models is needed to confirm these results.

The \FeiiF\ emitting region is more difficult to link to stable or radioactive material since most of the models predict that there is a non-negligible amount of stable $^{54}$Fe and $^{56}$Fe produced in the explosion. These contributions can not be easily disentangled. We have found that the \FeiiF\ and \NiIIF\ line have significant positive and negative velocity shifts with respect to the rest wavelengths. Interestingly, the shifts of the two features correlate, i.e.~when the \FeiiF\ emission lines are shifted to the blue, the \NiIIF\ lines usually do as well. The \FeiiF\ and \NiIIF\ velocity shifts are also significantly higher (up to $\pm$2000 \kms) than those of the \CoiiiF\ lines, which are consistent with zero shift.

A toy model was set up by \cite{2010ApJ...708.1703M} with i) an offset (from zero velocity) high-density region dominated by electron-capture elements (90 per cent by mass as stable $^{58}$Ni, 10 per cent of $^{56}$Fe), ii) an offset  relatively high-density  region with $^{56}$Fe, and iii) an extended, non-offset low-density region with $^{56}$Fe. They assumed in this model that the NIR lines of \FeiiF\ (12567 \AA, 16440 \AA) and the \FeiiF\ 7155 \AA\ line should come from the offset high-density $^{56}$Fe-dominated region.  Our result that the \CoiiiF\ lines come from a non-offset (zero velocity shift), extended (broader width)  $^{56}$Fe-dominated region agrees with the toy model of  \cite{2010ApJ...708.1703M}. The shifts to the red and blue for the \FeiiF\ and \NiIIF\ lines can be interpreted as viewing angle effects of an asymmetric explosion (see further discussion below in Section \ref{sec:asym_model}). 

\subsection{Asymmetry predictions from explosions models}
\label{sec:asym_model}

\cite{2016MNRAS.462.1039B} investigated the line velocities and polarisation signals of the DDT model (`N100') of \cite{2013MNRAS.429.1156S} and the sub-M$_{ch}$ double-detonation model of \cite{2010A&A...514A..53F}. They found that both models have continuum polarisation levels that match well with the observations but differ significantly in the level of asymmetries, with the sub-M$_{ch}$ double-detonation model having significantly higher levels of global asymmetry. These asymmetries are introduced by the explosion mechanism,  where the detonation is ignited at a single point on the surface and the detonation wave then travels on this surface to converge at the opposite side. This results in intermediate-mass elements that have a wider range of velocities and are more abundant on one side of the ejecta than on the other \citep{2010A&A...514A..53F}. The N100 DDT model of  \cite{2013MNRAS.429.1156S} does not predict significant variations in the offset of stable material (the stable $^{58}$Ni offset averaged over the ejecta is of the order of just tens of \kms) or variation in the \SiII\ velocity with viewing angle. For one of the most asymmetric models (N0003) of \cite{2013MNRAS.429.1156S}, the average velocity offset is larger but still not very large: $\sim$200 \kms. 

We have confirmed and strengthened the trend first identified by \cite{2010ApJ...708.1703M} that the SNe Ia with more redshifted velocities of the late-time \FeiiF\ lines are found to have higher maximum-light \SiII\ velocities (see Fig.~\ref{fig:vel_corr}). This suggests that the  \SiII\ velocity is greatest when the bulk of the Fe-group elements are moving away from the observer. This is in good qualitative agreement with the sub-M$_{ch}$ double-detonation model predictions, where an observer would see a high \SiII\ velocity when viewing the explosion from the opposite side of the ejecta to where the core detonation has occurred. The Fe material would in this case to be moving away from the observer \cite[see fig.~2 of][]{2016MNRAS.462.1039B}.  

 \cite{2015MNRAS.454L..61D} investigated the shape of the \CoiiiF\ 5890 \AA\ feature in a sample of 18 SNe Ia from the literature and found that some SNe Ia with $\Delta m_{15}(B)>1.3$ (stretch $<$0.85) have double-peaked profiles. They suggested this could be due to different viewing angles of a bimodal explosion of the direct collisions of white dwarfs. They suggested that \FeiiF\ lines should also show double-peaked features. We have not identified double-peaked features of \CoiiiF\ or \FeiiF\ in  the SNe Ia in our sample. Recent triple system population synthesis suggests that direct collisions are unlikely to explain more than 0.1 per cent of normal SNe Ia \citep{2017arXiv170900422T}.

\subsection{Constraints on Ni to Fe abundance}
Different explosion models predict different production rates of stable material due to different central densities at the time of explosion, with only M$_{ch}$ mass white dwarfs able to produce neutron-rich isotopes, such as $^{58}$Ni, during the explosion itself. However, sub-M$_{ch}$ models can have non-zero amounts of neutron-rich isotopes but this is dependent on the white dwarf metallicity that results in higher neutronisation for higher metallicity progenitors. The ratio of Ni to Fe is much lower for sub-M$_{ch}$ models than for M$_{ch}$ models, except for when very high metallicity (many times solar) sub-M$_{ch}$ progenitors are invoked \citep{2010ApJ...714L..52S, 2013MNRAS.429.1425R, 2018ApJ...854...52S}.

Stable Fe is difficult to measure at late times because radioactively produced $^{56}$Fe from the decay of $^{56}$Ni via $^{56}$Co is also present. However, at late times, all the radioactively-produced $^{56}$Ni has decayed and so any features of Ni seen in SN Ia spectra should be dominated by stable $^{58}$Ni. The most prominent lines to study are the doublet of \NiIIF\ 7378, 7412 \AA. We find that a contribution from \NiIIF\ is required to fit this region. 

We estimate the Ni/Fe abundance ratio to be in the range 0.02 -- 0.10 for the 7 SNe Ia studied here, within uncertainties on the values of $\sim$40 per cent, excluding systematic uncertainties, such as those due to choosing a certain set of atomic data. In Fig.~\ref{nife_abund}, we show that Ni/Fe abundance for our sample is broadly consistent with the M$_{ch}$ DDT models of \cite{2013MNRAS.429.1156S}. However, as discussed in Sec.~\ref{sec:asym_model}, the measured velocity offsets seen for the \FeiiF\ and \NiIIF\ lines appear inconsistent with this model. While there is overlap between the data and the sub-M$_{ch}$ explosion models of \cite{2010ApJ...714L..52S} and \cite{2018ApJ...854...52S}, they generally underproduce the Ni/Fe abundance compared to the observations and the sub-M$_{ch}$ models that match best are those with progenitor star metallicities significantly above solar. This increase in metallicity boosts the neutron excess available to produce stable Fe-group elements.  The `W7' model significantly over-predicts the Ni/Fe abundance for our sample, because of its large stable Ni core. Previous studies of SNe Ia have also found that the high nickel content of W7 leads to too strong emission lines of \NiIIF\ at late times \citep{1992ApJ...400..127R,1997ApJ...483L.107L,2009A&A...505..265L,2015ApJ...814L...2F}. 

To make a crude estimate of the number of SNe Ia occurring in super-solar metallicity environments, we used the results of a number of spectroscopic metallicity studies of the global host galaxy properties of SNe Ia. We find that the percentage of SNe Ia in high-metallicity galaxies is approximately half \citep[46 -- 64 per cent;][]{2013ApJ...770..107C,2014MNRAS.438.1391P,2016MNRAS.457.3470C}. However, to explain the majority of Ni over Fe abundances in our sample, a metallicity of $>$2 times solar would be required, which is reasonably rare for SNe Ia host galaxy locations -- estimates of the number of SNe Ia exploding in galaxies at greater than twice solar metallicity are of the order of $\sim$1--25 per cent \citep{2013ApJ...770..107C,2014MNRAS.438.1391P,2016MNRAS.457.3470C}. 

Studies of the local host galaxy properties of SNe Ia find similar numbers of events at greater than solar metallicity ($\sim$50 per cent), and at greater than twice solar metallicity ($\lesssim$5 per cent) \citep[Galbany et al. in prep.]{2016A&A...591A..48G}. We also found that our SN sample all occur in emission-line, morphologically spiral galaxies, suggesting that they are not biased towards very high-metallicity environments. Therefore, with the current sub-M$_{ch}$ detonation models it appears difficult to explain the bulk of the SN Ia Ni over Fe abundances observed in our late-time spectral sample.

Previous modelling efforts have suggested that \CaiiF\ emission could play a role in this region \citep{1997ApJ...483L.107L,2015ApJ...814L...2F,2015MNRAS.450.2631M,2017ApJ...845..176B}. The explosion models generally predict that intermediate-mass elements, such as Ca, have a shell-like ejecta structure. The presence of any broad \CaiiF\ emission would have a pedestal effect, increasing the level of the underlying continuum. We have found that a contribution is not needed in our analysis to fit the 7300 \AA\ region (Fig.~\ref{fig:feni}) and is rejected as a component in our fits using a model-selection technique. If we assumed that the redward peak in the 7300 \AA\ region had a contribution from Gaussian-shaped \CaiiF\ lines instead of \NiIIF\ lines then it would produce much larger velocity offsets, with the maximum redshifted SN 2012fr being shifted by  $\sim$5500 \kms\ -- a very large shift for any explosion model. Therefore, our conclusion is that \CaiiF\ emission does not significantly contribute to the 7300 \AA\ region of late-time SN Ia spectra.

The measured Ni/Fe abundance ratios for our sample of SNe Ia at late times are broadly consistent (mean of 0.06$\pm$0.03) with the solar Ni/Fe value of 0.057 \citep{2009ARA&A..47..481A}. A previous study of Ni/Fe abundances in a sample of six core-collapse SNe found significantly super-solar ratios in two of these objects, and around solar for the other four \citep{2015MNRAS.448.2482J}. It was speculated that SNe Ia may have, on average, lower than solar Ni/Fe ratios that would bring the combined yield of SNe Ia and core-collapse SNe to the solar value \citep{2015ApJ...807..110J}. However, we have shown that the Ni/Fe ratio for SNe Ia is broadly consistent with the solar value. Future studies of the Ni/Fe abundances in both core-collapse SNe and SNe Ia are needed so that we can put stronger constraints on their explosions, and their relative contributions to Galactic chemical evolution. 

\begin{figure}
\includegraphics[width=7.6cm]{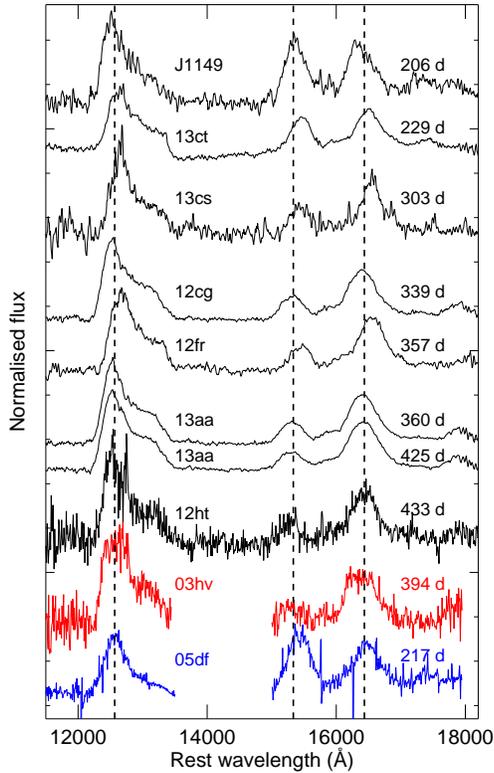} 
\caption{Comparison of SN Ia NIR spectra, focussing on the $\sim$11000-18000 \AA\ wavelength region. The 8 new XShooter spectra of 7 SNe Ia are shown in black compared to two from the literature (SN 2003hv in red and SN 2005df in blue). The dashed vertical lines marked the position of the most prominent \FeiiF\ emission features. No flat-topped profiles are identified for the XShooter spectral sample. Shifts of the peaks of the main \FeiiF\ features with respect to their rest wavelength can also be observed. In particular, shifts to the red for the peaks of the 12567 \AA\ and 16440 \AA\ lines can be seen for SN 2013ct, SN 2013cs, and SN 2012fr. }
\label{flat_tops}
\end{figure}

\subsection{Absence of flat-topped NIR \FeiiF\ features}

`Flat-topped' profiles have been suggested to arise in SNe Ia originating from M$_{ch}$ explosions, where the high central density at the time of explosion results in significant electron capture resulting in a core that is dominated by stable Fe-group elements \citep{2004ApJ...617.1258H}. At $>$150--200 d, local deposition by positrons should dominate over gamma-ray deposition \cite[e.g.][]{2014ApJ...795...84P}. Therefore, if there is no mixing of the ejecta, this highest density region could produce nebular-phase emission lines of radioactivity-dominated species that have flat-topped profiles (due to the lack of radioactive elements in this region). Previous studies have identified flat-topped profiles at NIR wavelengths for SNe 2003du and 2003hv \citep{2004ApJ...617.1258H,2006ApJ...652L.101M}. They were not seen at optical wavelengths, which was suggested to be due to more line blending at optical wavelengths. Flat-topped profiles were not seen in the NIR for SN 2005df \citep{2015ApJ...806..107D} or SN 2011fe \citep{2015MNRAS.450.2631M}. 

For our XShooter sample, we have found that the 12600 \AA\ region is well fit by multiple Gaussians of \FeiiF\ lines (see Fig.~\ref{fe_126m}). The main peak of this feature comes nearly exclusively from the \FeiiF\ 12567 \AA\ line and we identified no evidence for a flat-topped profile of this or the other weaker \FeiiF\ features in this region. In Fig.~\ref{flat_tops}, we show the NIR region from $\sim$11000--18000 \AA\ for our sample of 8 late-time SN Ia spectra, as well as the $+$394~d spectrum of SN 2003hv \citep{2006ApJ...652L.101M,2009A&A...505..265L} and $+$217 d spectrum of SN 2005df \citep{2015ApJ...806..107D}. As previously discussed, we did not include the 15000--18000 \AA\ region in our detailed fitting analysis due to the more complicated temperature dependence of \FeiiF\ lines in this region and potential contributions from other elements. However, from a qualitative analysis of Fig.~\ref{flat_tops}, we do not identify flat-topped profiles for the SNe in our sample in any of the main \FeiiF\ emission features in the NIR. Although it cannot be ruled out that the previously identified flat-topped profiles are real, it is possible that they were due to low S/N in these regions.

We also do not require flat-topped profiles for the \FeiiF\ at optical wavelengths, which are predicted by models that only have stable material in their highest density regions \citep{2004ApJ...617.1258H}. A central region solely comprised of stable material is not required for our spectra. The widths of the \NiIIF\ lines of $\sim$5000 -- 9000 \kms\ also overlap with those of the \FeiiF\ features of $\sim$7000 --9000 \kms, again suggesting that the stable material that we have identified in our spectra is not confined to a high-density region without any radioactive material present.

\begin{figure}
\includegraphics[width=8.8cm]{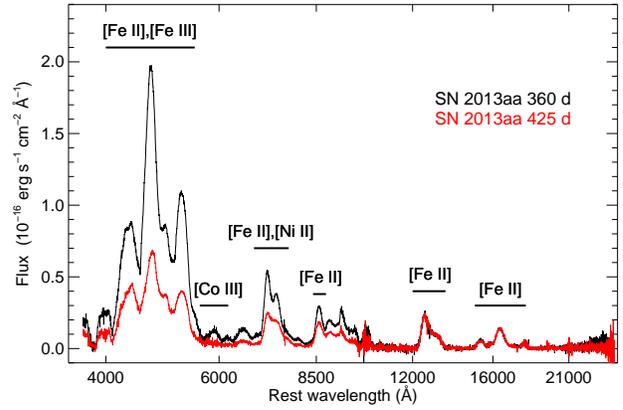} 
\caption{Comparison of the optical and NIR spectra of SN 2013aa at 360 and 425 d post maximum. The main contributors to individual spectral regions are marked.}
\label{2013aa_comp}
\end{figure}
\subsection{Cooling at late-times}
As has been seen in other studies of late-time data of SNe Ia, there is a marked increase in flux emerging at NIR compared to optical wavelengths at $>$350 d past explosion (Fig.~\ref{nir_percent}). This is seen for SN 2013aa in Fig.~\ref{2013aa_comp}, which has spectra at 360  and 425 d post maximum. The flux in the NIR \FeiiF\ lines remains constant between the two epochs but the flux at bluer wavelengths has significantly decreased. This levelling off in the flux of the NIR bands of SNe Ia at late times was previously seen in SN 1998bu  \citep{2004A&A...426..547S} and SN 2000cx \citep{2004A&A...428..555S} in the phase range of $\sim$250 -- 480~d. Since most of the emission at these phases is still from thermal collisions, this relative increase in the the flux in the NIR bands is expected to be due to the continued cooling of the ejecta, resulting in the transition to emission at longer wavelengths. This temporarily cancels out the decreasing radioactive heating at NIR wavelengths resulting in little evolution in the NIR bands at these epochs.

We have also identified a hint of the transition from \FeiiiF/\FeiiF\ to \FeiiF/\FeiF-dominated spectra that was previously observed in SN 2011fe between 365 and 593 days post explosion \citep{2015MNRAS.454.1948G}. The main peak in the blue at $\sim$4700 \AA\ is dominated by \FeiiiF\ emission while the peak to the red of this at $\sim$5300 \AA\ is mixture of \FeiiiF\ and \FeiiF. If we normalise to this peak at 5300 \AA\ we find that the peak at 4700 \AA\ has decreased with respect to it, meaning that \FeiiiF/\FeiiF\ ratio has decreased between theses epochs. We have not identified any \FeiF\ emission features in our spectra but speculate based on the results of \cite{2015MNRAS.454.1948G} that we may be seeing the beginning of a transition from an  \FeiiiF/\FeiiF\ to \FeiiF/\FeiF-dominated spectrum in the range of 360 and 425 d post maximum for SN 2013aa.

\section{Conclusions}
\label{sec:conc}

In this paper, we have presented a detailed analysis of the late-time ($>$200 d post explosion) optical and NIR spectra of 7 SNe Ia. The late-time XShooter spectra were supplemented by maximum-light photometric and spectroscopic observations. Using a multi-Gaussian fitting code, we have performed fits to three prominent spectral regions: the 12600 \AA\ \FeiiF-dominated region, the 7300 \AA\ \FeiiF/\NiIIF-dominated region, and the 6000 \AA\ \CoiiiF\ region. This has allowed us to constrain the stable Ni over Fe abundance ratio, the velocity shifts and widths of the measured features, and their connection to maximum-light observations. We have investigated the NIR compared to optical flux contribution, along with a study of the presence of flat-topped NIR profiles and the double-peaked \CoiiiF\ features.  

Our main results are:
\begin{enumerate}
\item Multi-Gaussian fits to the 7300 \AA\  \FeiiF/\NiIIF-dominated region show that a contribution from stable \NiIIF\ is required to fit the line profiles (Figs.~\ref{fig:feni} and \ref{fig:feni_noni}). \CaiiF\ is found to make a negligible contribution to this region.
\item The abundance ratio of Ni to Fe is found to be in the range of 0.02 -- 0.10 for the 7 SNe Ia in our sample, with uncertainties of $\sim$40 per cent, making them broadly consistent with DDT and some super-solar sub-M$_{ch}$ models. The mean value of our sample is consistent with solar Ni/Fe abundance measurements.
\item Using a simple fit to the 4700 \AA\ \FeiiiF\ main peak, we confirm the results of previous studies that the peak of this feature shifts to the red with time.
\item We have constrained the velocity offsets (from rest wavelength) of the \FeiiF, \NiIIF, and \CoiiiF\ features. The \FeiiF\ 7155 \AA, 12567 \AA\ and the \NiIIF\ 7378 \AA\ features are found to have significant offsets ($\sim$500--2000 \kms) both positive and negative from their rest wavelengths (Fig.~\ref{fig:vel_width_shift}). The \CoiiiF\ 5890 \AA\ feature does not show such significant offsets with a mean absolute velocity offset of 271$\pm$341 \kms, suggesting a different emitting region for the \CoiiiF\ compared to the \FeiiF\ and \NiIIF\ features.
\item The \CoiiiF\ 5890 \AA\ feature is found to have significantly broader FWHM of $\sim$10000 \kms\ than the \FeiiF\ and \NiIIF\ features of 6600 -- 7600 \kms, suggesting a broader emitting region for \CoiiiF\ compared to \FeiiF\ and \NiIIF\ features. 
\item By connecting the late-time \FeiiF\ line shift velocities with the velocity of the \SiII\ absorption feature at maximum light, we confirm (and strengthen) the trend of \cite{2010Natur.466...82M} where SNe Ia with blue-shifted \FeiiF\ features have lower \SiII\ velocities. 
\item A trend of increasing NIR flux with time at epochs $>$350 d past explosion is identified (from $<$10 at $\sim$350 d up to 40 per cent at $\sim$550 d). The onset of the transformation from \FeiiiF/\FeiiF- to \FeiiF/\FeiF-dominated spectra is potentially identified for SN 2013aa in the range of 360 -- 425 d post maximum light.
\item In contradiction with some previous studies, we do not see flat-topped profiles for the \FeiiF\ features at NIR wavelengths, which may be due to the higher S/N spectra in our sample.
\item No double-peaked \CoiiiF\ or \FeiiF\ features are identified that have been suggested to be present in fainter SNe Ia in direct collision models.
\end{enumerate}

The largest sample to date of late-time combined optical and NIR spectra of SNe Ia has allowed us to test a  number of theories related to density structures, asymmetries, and elemental abundances. We have identified signatures of potential asymmetries in their explosions through measurements of the velocities and widths of key Fe-group elements. Our results suggest that the explosions of normal SNe Ia, as studied here, may be intrinsically non-spherically symmetric with significant offsets from zero velocity for some elements and ionisation states. Future work will involve larger samples of SNe Ia at late times, combined with state-of-the-art spectral modelling (Shingles et al.~in prep.), to allow the geometry and explosion products to be quantified to higher precision, and place tighter constraints on viable SN Ia explosion models.

\section{Acknowledgements}
 KM is supported by STFC through an Ernest Rutherford Fellowship (ST/M005348/1). SS, LS, and KM acknowledge support from STFC through grant, ST/P000312/1.   TWC is supported by an Alexander von Humboldt Fellowship.  LG was supported in part by the US National Science Foundation under grant AST-1311862.  CPG acknowledges support from EU/FP7-ERC grant No. [615929]. DAH and GH are supported by NSF AST-1313484.  RR received support from The Aerospace CorporationÕs Technical Investment Program. Based on data taken at the European Organisation for Astronomical Research in the Southern Hemisphere, Chile, under program IDs: 091.D-0764(A), 092.D-0632(A), 096.D-0627(A), and as part of PESSTO (188.D-3003). This research has made use of the NASA/IPAC Extragalactic Database (NED) which is operated by the Jet Propulsion Laboratory, California Institute of Technology, under contract with the National Aeronautics and Space Administration.  This work makes use of the LCO network. 

The Pan-STARRS1 Surveys (PS1) and the PS1 public science archive have been made possible through contributions by the Institute for Astronomy, the University of Hawaii, the Pan-STARRS Project Office, the Max-Planck Society and its participating institutes, the Max Planck Institute for Astronomy, Heidelberg and the Max Planck Institute for Extraterrestrial Physics, Garching, The Johns Hopkins University, Durham University, the University of Edinburgh, the Queen's University Belfast, the Harvard-Smithsonian Center for Astrophysics, the Las Cumbres Observatory Global Telescope Network Incorporated, the National Central University of Taiwan, the Space Telescope Science Institute, the National Aeronautics and Space Administration under Grant No. NNX08AR22G issued through the Planetary Science Division of the NASA Science Mission Directorate, the National Science Foundation Grant No. AST-1238877, the University of Maryland, Eotvos Lorand University (ELTE), the Los Alamos National Laboratory, and the Gordon and Betty Moore Foundation.

\bibliographystyle{mn2e}
\bibliography{astro}

\newpage

\appendix \label{appa}
\section{Photometry of SN 2013cs}
\begin{table}
 \caption{Photometry for SN 2013cs obtained with the LCO 1-m array.}
 \label{tab:sn13cs_phot}
\begin{tabular}{@{}lccccccccccccccccccccccccccccc}
  \hline
  \hline
MJD$^a$&\textit{g}&\textit{r}&\textit{i}\\
&(mag)&(mag)&(mag)\\
\hline
\hline
56427.1  &  14.82$\pm$0.06   &14.74$\pm$0.08    &15.17$\pm$0.11 \\
56427.7  &  14.71$\pm$0.05   &14.63$\pm$0.06    &15.04$\pm$0.17 \\
56428.8  &  14.50$\pm$0.05   &14.47$\pm$0.07    &14.80$\pm$0.12  \\
56431.1  &  14.15$\pm$0.04   &14.14$\pm$0.04    &  -- \\
56448.0  &  14.45$\pm$0.07   &14.48$\pm$0.09    &15.27$\pm$0.06  \\
56453.7  &  14.82$\pm$0.04   &14.59$\pm$0.06    &15.43$\pm$0.20  \\
56454.0  &  14.91$\pm$0.05   &14.64$\pm$0.06    &15.31$\pm$0.12  \\
56459.0  &  15.29$\pm$0.06   & --			     &15.18$\pm$0.06  \\
56461.0  &  15.43$\pm$0.06   &14.92$\pm$0.06    &  -- \\
56463.0  &  15.63$\pm$0.05   &14.58$\pm$0.08    &15.13$\pm$0.09  \\
56465.0  &  15.70$\pm$0.05   &14.82$\pm$0.06    &15.00$\pm$0.10  \\
56467.0  &  15.99$\pm$0.04   &14.99$\pm$0.06    &15.17$\pm$0.13  \\
56469.0  &  16.22$\pm$0.05   &15.16$\pm$0.08    &15.31$\pm$0.09 \\
\hline
\end{tabular}
 \begin{flushleft}
$^a$MJD = Modified Julian date.\\
  \end{flushleft}
\end{table}

\end{document}